\documentclass[aps,prd,preprint,onecolumn,superscriptaddress,tightenlines,nofootinbib,eqsecnum,preprintnumbers,longbibliography,12pt]{revtex4-1}
\pdfoutput=1
\usepackage{amsmath,latexsym,amssymb,graphicx,ifpdf, slashed, color,url, cancel, comment}
\usepackage[T1]{fontenc}
\usepackage{relsize}
\usepackage{ulem}
\usepackage{braket}
\usepackage{enumitem}
\usepackage[x11names,dvipsnames]{xcolor}
\usepackage{hyperref}

\definecolor{niceblue}{rgb}{0.1, 0.3, 0.8}
\definecolor{BrickRed}{rgb}{0.8, 0.25, 0.33}
\hypersetup{colorlinks,citecolor= BrickRed, linkcolor= niceblue, urlcolor=niceblue}

\begin{document}
\def\MPIK{ICTP, Italy}

\preprint{}

\title{Refractive Neutrino Mass: \\Effects of Fluctuations of the Dark Matter Field}

\author{Soham Sahasrabuddhe}
\email{sohams@iitb.ac.in}
\author{Manibrata Sen}
\email{manibrata@iitb.ac.in}
\affiliation{Department of Physics, Indian Institute of Technology Bombay, Powai, Mumbai 400076, India}
\author{Alexei Y. Smirnov}
\email{smirnov@ictp.it}
\affiliation{International Centre for Theoretical Physics, ICTP, Strada Costiera 11, 34151 Trieste, Italy}

\date{\today}

\begin{abstract}

Observed neutrino oscillations can be reproduced by a refractive neutrino mass-squared arising from interactions with ultralight scalar dark matter (ULDM). 
The wave behaviour of the scalar field gives rise to a fluctuating interference pattern, inducing spatial and temporal variations in the refractive mass that can distort neutrino oscillation probabilities. We simulate the interference patterns of the scalar field in the presence of self-interactions. Using results of these simulations, the oscillation probabilities for JUNO and T2K experiments are computed. We find that the refractive mass scenario remains viable for the complex scalar field without self-coupling for $m_\phi \lesssim 10^{-16}~\mathrm{eV}$. 
For $m_\phi\gtrsim10^{-16}~\mathrm{eV}$, repulsive self-interactions narrow the density distribution, reducing the deviation of the survival probability at the first minimum to $\Delta P \approx 0.07$ for an appropriate self-interaction coupling.
In contrast, for a real scalar field, rapid oscillations with the Compton frequency distort the probabilities for scalar masses $ 10^ {- 22}~\mathrm{eV} \lesssim m_\phi \lesssim 10^{-10} ~\mathrm{eV}$, and self-interactions do not remove this distortion. This distortion becomes negligible only for $m_\phi\gtrsim10^{-9}~\mathrm{eV}$. We explore additional effects which could modify the interference pattern and relax the constraints on the required values of $m_\phi$. These include scenarios with multiple scalar fields (the axiverse), the gravitational effects of the Sun and Earth on the granule structure, and interactions with the relic neutrino background.

\end{abstract}

\maketitle

\section{Introduction}

The identification of the origins of neutrino mass and its smallness remains a key long-standing open issue. Neutrino experiments, particularly those probing oscillations, are sensitive to the neutrino mass squared, $|m|^2$. Since neutrino oscillations do not involve a chirality flip, interactions that preserve chirality (e.g., vector-like interactions) or involve two chirality flips (e.g., two scalar interactions) can give rise to an effective neutrino mass squared. This observation motivates the introduction of the refractive neutrino mass squared generated by neutrino forward scattering on particles in a medium~\cite{Choi:2019zxy, Choi:2020ydp, Chun:2021ief, Sen:2023uga}. In particular, Ref.~\cite{Sen:2023uga} considered neutrino scattering on very ultralight scalar dark matter (ULDM) particles $\phi$, mediated by a very light fermion $\chi$, a singlet under the Standard Model gauge group.
The corresponding interaction term in the effective Lagrangian is
\begin{equation}
\mathcal{L} = g_\phi \bar{\nu} \chi \phi + \mathrm {h.c.}\,
\label{eq:lagran}
\end{equation}
For ultra-relativistic neutrinos, the mass-dependent term in the evolution Hamiltonian is inversely proportional to the neutrino energy: $|m|^2/2E$. For this reason, the original proposal by Wolfenstein of inducing oscillations of massless neutrinos through vector-like interactions \cite{Wolfenstein:1977ue} does not work: at energies below the mediator mass, the potential $V$, and consequently, the oscillation probabilities, do not depend on energy, in contrast to observations.

The situation is different when the mediator, $\chi$, is very light.
Above the resonance energy, determined by the mediator and the ULDM masses, $E_R \equiv m_\chi^2/ 2 m_\phi$~\cite{Smirnov:2021zgn}, the potential scales inversely with the neutrino energy, just like the mass-dependent term in the evolution Hamiltonian. Consequently, the quantity $(2VE)$ becomes energy-independent and can serve as an effective mass squared.

Reproducing the observed oscillation effects requires a sufficiently large number density of scattering particles~\cite{Sen:2023uga}. For scalar particles coupled to neutrinos with strength $g_\phi$, as given by Eq.~\eqref{eq:lagran}, we have
\begin{equation}
g_\phi^2 n_\phi/m_\phi = \rho_\phi g_\phi^2/m_\phi^2 
\sim \Delta m^2\,,
\label{eq:refaction}
\end{equation}
where $\rho_\phi$ is the energy density of the scalar particles and $\Delta m^2$ denotes the mass-squared differences measured in neutrino oscillation experiments. Eq.~\eqref{eq:refaction} implies that the scalar particles must constitute all or a substantial fraction of the dark matter (DM) in the Universe. 
If $\rho_\phi = \rho_{\rm DM}$, reproducing the observed neutrino mass-squared differences requires the scalar parameters to satisfy 
\begin{equation}
\frac{g_\phi}{m_\phi} \simeq (5-30) \, {\rm eV}^{-1}
\sqrt{\frac{\rho_{\rm DM}}{\rho_\phi}}.
\label{eq:strength}
\end{equation}

Various constraints apply to the scalar mass and its interactions with neutrinos. Ref.~\cite{Sen:2023uga} identifies two allowed mass regions:
\begin{equation}
m_\phi < 10^{-19}\, {\rm eV} \quad {\rm and} \quad m_\phi = \left(10^{-12} - 10^{- 8}\right)\, {\rm eV}. 
\label{eq:allowed-m}
\end{equation}
The corresponding values of the coupling constant and mediator mass
($\chi$ is a light singlet fermion/sterile neutrino) are
\begin{equation}
g_\phi = (5 - 30) \,\frac{m_\phi}{{\rm eV}}\quad {\rm and} \quad m_\chi = 10^{-3} 
\, {\rm eV} \, \sqrt{\frac{m_\phi}{10^{-9} \rm eV}}. 
\label{eq:pheno}
\end{equation}
In what follows, we use these allowed regions as benchmark scenarios. 
Additional new physics could enlarge the allowed parameter space. 
Our analysis therefore considers the scalar mass range $10^{-22}\,{\rm eV}\lesssim m_\phi\lesssim 10^{-8}\,{\rm eV}$.

A key feature of the refractive mass is its dependence on the number density of scatterers (DM particles), which makes it vary with both time and position. In particular, the effective mass squared is larger in the central regions of the galaxy, potentially allowing the refractive mass scenario to be probed through time delays in supernova (SN) neutrino signals from explosions at different galactic locations~\cite{Ge:2024ftz}. 
The delay is expected to be larger for SNe occurring closer to the galactic centre, or if there exists a DM spike along the neutrino path~\cite{Pompa:2025lbf}. This spatial dependence also weakens the constraint on the refractive mass from SN1987A, since the neutrinos propagated predominantly through regions of relatively low DM density and therefore experienced a smaller effective mass.

The refractive mass also depends on energy below the resonance energy $E_R$, decreasing with energy and becoming very small for relic neutrinos. This energy dependence modifies the cosmological implications, particularly the constraint on the sum of neutrino masses from structure formation~\cite{Sen:2024pgb}. The effective neutrino mass squared in intergalactic space is five orders of magnitude smaller than that in the galaxy, although it was larger in the past.

Refractive neutrino mass requires the DM to be ultralight. Owing to the high number density 
of the scattering particles, the DM can be treated as a classical scalar field. 
In earlier studies, the scalar field was considered to be spatially uniform 
and constant over the duration of observations or experiments. In reality, 
the field exhibits wave-like behaviour, giving rise to an interference pattern 
with highly non-uniform and stochastic variations in space and time. 
The galactic spacetime pattern consists of a stable solitonic core surrounded 
by a halo with fluctuating granules and 
voids~\cite{Schive:2014dra, Hui:2016ltb, Boehmer:2007um, Hui:2020hbq, Hui:2021tkt, Schive:2025bcm, Hu:2000ke}. 
Most previous studies applying the scalar-field framework to neutrino physics have focused 
on temporal variations associated with the dominant oscillation mode, 
whose frequency is set by the scalar mass $m_\phi$~\cite{Berlin:2016woy, Krnjaic:2017zlz, Brdar:2017kbt, Liao:2018byh, Capozzi:2018bps, Ge:2018uhz, Dev:2020kgz, Losada:2021bxx, Dev:2022bae, ChoeJo:2023ffp, Losada:2022uvr, Losada:2023zap,Ge:2019tdi, Huang:2021kam, Huang:2022wmz, Davoudiasl:2023uiq, Sahu:2025vyy}. The refractive nature of neutrino mass has also recently been used to link quark and leptonic mixing, proposing that leptonic mixing arises from DM medium effects.~\cite{Chattopadhyay:2026wil}.

The effects of field fluctuations arising from wave interference on neutrino oscillations were investigated in~\cite{Cheek:2025kks}. 
They found that, for $m_\phi \gg 10^{-14}\text{ eV}$, the de Broglie wavelength of the scalar field becomes 
smaller than the distance traversed by the Earth during a typical experimental exposure. Consequently, 
the Earth crosses regions with varying DM densities, and the resulting spacetime averaging over both rapid field oscillations and stochastic density 
fluctuations suppresses the oscillatory behaviour of the neutrino oscillation probability 
as a function of propagation distance and energy, ruling out this mass regime in a model-independent manner. 
For $3 \times 10^{-19}\text{ eV} \lesssim m_\phi \lesssim 10^{-14}\text{ eV}$, the field amplitude remains approximately spatially uniform over the experimental baseline, but the rapid temporal oscillations of the real scalar field still distort neutrino propagation. 
After averaging the oscillation probabilities over these temporal variations, a comparison with KamLAND measurements of the $\bar{\nu}_e$ survival probability as a function of $L/E$ disfavours the dark-origin hypothesis for neutrino masses at the $4.5\sigma$ level.

Ref.~\cite{Chathirathas:2026bgu} refined the analysis of~
\cite{Cheek:2025kks} and extended it to the case of a complex scalar field. The effects were evaluated using a stochastic model of the scalar field, without performing numerical simulations. In this model, the field is represented as a superposition of plane waves with different momenta. Their amplitudes are weighted according to a Maxwellian velocity distribution and multiplied by random factors drawn from a Rayleigh distribution. Additionally, the waves also carry independent random phases for different momenta. The resulting statistical distributions are independent of $m_\phi$ and other dimensional parameters, 
allowing the model to be applied to different scalar masses by appropriately rescaling the spatial and temporal scales. For a real scalar field, Ref.~\cite{Chathirathas:2026bgu} found that, over the range $10^{-21} \, {\rm eV} \lesssim m_\phi \lesssim 3 \times 10^{-8} \, {\rm eV}$, only approximately 9\% to 54\% of the neutrino mass can be of refractive origin. In contrast, for the case of a complex scalar field, the oscillation probabilities remain undistorted for $m_\phi < 10^{-16}$ eV, leaving the refractive mass scenario viable in this range.

In this paper, in contrast to Refs.~\cite{Chathirathas:2026bgu, Cheek:2025kks} we consider neutrino refraction and oscillations in the scalar field with self-interactions, which are generated anyway at the loop level by the coupling in Eq.~\eqref{eq:lagran}. 
We find that self-interactions can significantly suppress the distortions of neutrino oscillation probabilities induced by scalar-field fluctuations. 
Moreover, additional effects, including the presence of multiple scalar species and solar and terrestrial gravitational fields, can further weaken the constraints, leaving refraction-induced neutrino masses viable over a broad range of scalar masses.

The paper is organised as follows. In Sec.~\ref{sec:Dmf}, we discuss the key features of the scalar DM interference pattern and present the results of our simulations. In Sec.~\ref{sec:refrac}, we examine how variations in the scalar field affect neutrino oscillations and calculate the oscillation probabilities for different scalar masses and self-interaction strengths. In Sec.~\ref{sec:modif}, we consider modifications to the scalar field due to the effects of an external gravitational field and the presence of multiple scalar fields. We estimate how these modifications affect the oscillation probabilities and the constraints on the refractive neutrino mass. Finally, we present our conclusions in Sec.~\ref {sec:conclusion}.

\section{Dark matter field and interference pattern}
\label{sec:Dmf}

The interplay of gravity and ``quantum pressure'' associated with the large de Broglie wavelength of ULDM leads to an interference pattern characterised by highly non-uniform and stochastic variations in the field density~\cite{Schive:2014dra, Hui:2016ltb, Boehmer:2007um, Hui:2020hbq, Hui:2021tkt, Schive:2025bcm}. Gravity leads to the virialisation of DM particles, establishing a characteristic velocity distribution. The interference of waves with different velocities (momenta) produces a pattern with a stable solitonic core surrounded by halos of stochastic (fluctuating) granules and voids.

This interference pattern arises for non-relativistic DM particles even in the simplest scenario, where gravity is the only relevant interaction. Here, we present the results of our simulations that investigate how scalar self-interactions modify this interference pattern.
In our framework, such self-interactions can be introduced at tree level and can also be generated radiatively through loop diagrams involving $\nu$ and $\chi$ due to the interaction in Eq.~\eqref{eq:lagran}. It turns out that the loop-generated quartic self-interaction coupling is approximately
\begin{equation}
 \lambda \simeq \frac{g_\phi^4}{16\pi^2} \log \frac{q^2}{m_\chi^2}\,,
\label{eq:lambdaeqn}
\end{equation}
where $q$ is a reference four-momentum. This implies that the quartic coupling can lie within a range that significantly modifies the interference pattern.

\subsection{Evolution equation and its scaling properties}

We consider a complex scalar field for our analysis. The classical action for an ULDM field with quartic self-interactions, minimally coupled to gravity, is given by
\begin{equation}
S = \int d^4 x \sqrt{-g} \left[ -g^{\mu \nu} {\partial}_\mu \phi^* {\partial}_\nu \phi - 
m_\phi^2 |\phi|^2 - \lambda|\phi|^4 \right], 
\label{eq:action}
\end{equation}
where $m_\phi$ is the scalar mass and $\lambda$ is the dimensionless self-interaction coupling. The sign of $\lambda$ fixes the character of the interaction: $\lambda > 0$ is repulsive, $\lambda < 0$ is attractive, and $\lambda = 0$ is the pure gravitational regime. Here, the interaction in Eq.~\eqref{eq:lagran} is not included and will be treated as a small perturbation.

For a non-relativistic scalar field, the dominant rapid temporal oscillation occurs at the Compton frequency, $m_\phi$. We factor out this rapid oscillation from a slowly varying complex function $\psi$:
\begin{equation}
\phi = \frac{1}{\sqrt{2}\,m_\phi} \psi \, e^{-i m_\phi t }.
\label{eq:complexfield}
\end{equation}
The prefactor $1/(\sqrt{2}\, m_\phi)$ is chosen such that $|\psi|^2 = \rho_\phi$ 
is the energy density of the ULDM field. For a complex field, 
the components $\phi_-$ and $\phi_+$ which correspond to particles and antiparticles, 
do not interfere. This can be supported by the exact $U(1)$-symmetry of the underlying model, 
related e.g. to transformations $\phi \rightarrow e^{i \gamma} \phi$, 
$\phi^\dagger \rightarrow e^{-i \gamma} \phi^\dagger$, 
$\nu \rightarrow e^{i \gamma} \nu$ with $\chi$ being invariant. 
To keep the symmetry intact, $\phi$ should not acquire a vacuum expectation value (vev). Consequently, $|\phi|^2$ does not oscillate with frequency $m_\phi$.

In terms of particles, in the real field case, 
the $\nu \rightarrow \chi$ transition can be due to emission and absorption of $\phi$, 
while in the complex field case, that can be due to absorption of $\phi$ 
or emission of $\phi^\dagger$, which can not interfere. 
For this reason, the refraction effect above the resonance energy 
$E_R$ is proportional to $|\phi_+|^2 + |\phi_-|^2 = \rho_+ + \rho_-$. 
In contrast, at energies below the resonance, the amplitudes of scattering have opposite signs, 
and two terms can cancel each other, leading to zero refraction at the tree level \cite{Sen:2023uga}.
Thus, to have refractive mass, the asymmetry of the field $\phi$ is not needed. 
In general, the energy densities of particles $\phi_-$ and antiparticles $\phi_+$ 
are different and, in what follows, we will treat $\rho$ as the total energy density of scalars.
These conclusions in terms of particles at 
the quantum level should not change at the level of effective description in terms of classical fields.

With $\Phi_N$ as the Newtonian gravitational potential and $\bar{\rho}_\phi$ 
as the mean ULDM energy (mass) density, this leads to the Gross-Pitaevskii-Poisson (GPP) equations for the field $\psi$ 
\cite{Boehmer:2007um}:
\begin{align}
i {\partial}_t \psi & = -\frac{1}{2m_\phi} \nabla^2 \psi + 
m_\phi \Phi_N \psi + \frac{\lambda}{2 m_\phi^3} |\psi|^2 \psi, 
\label{eq:GPP}\\
\nabla^2 \Phi_N & = 4\pi G_N \left(|\psi|^2 -\bar{\rho}_\phi \right). 
\label{eq:GPP2}
\end{align}

The galactic dynamics determine characteristic halo size and velocity distribution, 
which are approximately independent of the scalar mass. 
The corresponding momentum dispersion scales as $k = \bar{v} m_\phi$, 
where $\bar{v} = 220 \, {\rm km/s}$ characterises the velocity dispersion, so increasing the scalar mass leads to a larger momentum spread.

In Eq.~\eqref{eq:GPP}, the $\nabla^2$ term represents the kinetic energy 
of the field. In momentum space, the Laplacian is proportional to $k^2$, and the evolution operator on the right-hand side of Eq.~\eqref{eq:GPP} can be viewed as the sum of the kinetic (first), gravitational (second), and self-interaction (last) contributions.

The de Broglie wavelength determines the spatial scale of the interference pattern,
\begin{equation}
\lambda_{\rm dB} =\frac{2\pi}{k} = \frac{2\pi}{m_\phi \bar v} =
1.7 \times 10^{9} {\rm cm} \left(\frac{10^{-10} {\rm eV}}{m_\phi}\right).
\label{eq:ldB}
\end{equation}
It gives the size of the granule,
distance between granules, and the coherence length over which the field oscillates with the same phase $(m_\phi t)$. It also characterises the scale at which quantum pressure becomes important. 

Let us introduce the following rescalings: 
\begin{equation}
x = \frac{1}{m_\phi \bar{v}} \tilde{x}, \, \, ~~
t = \frac{1}{m_\phi \bar{v}^2} \tilde{t}, \, \, ~~
\psi = \sqrt{\bar{\rho}_\phi}\tilde{\psi}, \, \, \, \, 
\Phi = \bar{v}^2 \tilde{\Phi}\,,
\label{eq:relations}
\end{equation}
 where the dimensionless coordinates $\tilde{x}$ and $\tilde{t}$ are the rescaled units. From Eq.~\eqref{eq:relations}, we have
$$
\tilde{x} = \frac{2\pi x}{\lambda_{\rm dB}}, \, \, \qquad
\tilde{t} = \frac{2 \pi t}{t_{\rm coh}}, 
$$
hence, $\tilde{x}$ is the distance in units of $\lambda_{\rm dB}/2\pi$, and $\tilde{t}$ is the time in units of $t_{\rm coh}/2\pi$. Substitution of these relations into Eq.~\eqref{eq:GPP} gives the dimensionless GPP equations: 
\begin{align}
i {\partial}_{\tilde{t}} \tilde{\psi} & = 
-\frac{1}{2} \tilde{\nabla}^2 
\tilde{\psi} + \tilde{\Phi} \tilde{\psi} + \beta|\tilde{\psi}|^2 \tilde{\psi}, 
\nonumber\\
\tilde{\nabla}^2 \tilde{\Phi} & = \alpha \left( |\tilde{\psi}|^2 - 1 \right), 
\label{eq:tildeGPP}
\end{align}
where 
\begin{equation}
\alpha \equiv \frac{4\pi G_N \bar{\rho}_\phi}{m_\phi^2 \bar{v}^4}, \, \qquad
\beta \equiv \frac{\lambda \bar{\rho}_\phi}{2m_\phi^4 \bar{v}^2}, 
\label{eq:relations2}
\end{equation}
characterise, respectively, the strength of gravity and the scale of self-interaction 
relative to the kinetic energy scale. Following Eq.~\eqref{eq:relations}, we get 
$\tilde{\psi} = \psi/ \sqrt{\bar \rho_\phi}$, therefore $\braket{|\tilde{\psi}|^2} = \braket{|\psi|^2}/ \bar{\rho}_\phi = 1$.

Notice that the parameter $\beta $ is constant if $\lambda \propto m_\phi^4$. 
According to Eq.~\eqref{eq:lambdaeqn}, this scaling is obtained when $g_\phi \propto m_\phi$, 
provided the logarithmic factor is approximately constant. 
This is the scaling required for self-interactions to have a substantial effect on the interference pattern (see Eq.~\eqref{eq:pheno}), so for couplings of interest, $\beta$ is independent of $m_\phi$.
In contrast, for fixed $\bar \rho_\phi$ and $\bar v$, 
we have $\alpha \propto m_\phi^{-2}$, so the gravitational term cannot be scaled out. 
Taking $\bar \rho_\phi = 0.4 \, {\rm GeV \, cm}^{-3}$, we find $\alpha = 8.9 \times 10^{-5}$ 
for $m_\phi = 10^{-22}$ eV and $\alpha =8.9 \times 10^{-25}$ for $m_\phi = 10^{-12}$ eV. 
Since $\alpha \ll 1$ throughout the mass range considered, the gravitational contribution is subdominant 
on the scales resolved by our simulations. Accordingly, the interference patterns remain essentially unchanged across this range.

If $\alpha$ can be neglected, and $\beta$ is independent of $m_\phi$, the system of 
equations Eq.~\eqref{eq:tildeGPP} exhibits an exact scaling with respect to $m_\phi$. 
A solution obtained in rescaled units for one value of $m_\phi$ therefore applies 
to any other value, provided the dimensionless parameters and initial conditions are unchanged. 
The dependence on $m_\phi$ is recovered when the solution is expressed in dimensional
units according to Eq.~\eqref{eq:relations}. We therefore use $m_\phi$ and $\beta$ as the input parameters to the calculation.

The characteristic timescale associated with the field dynamics is the coherence (kinetic) timescale, given as 
\begin{equation}
t_{\rm coh}  = 
 \frac{\lambda_{\rm dB}}{\bar{v}} = \frac{2\pi}{m_\phi \bar{v}^2}
= 77\, {\rm s} \left(\frac{10^{-10} {\rm eV}}{m_\phi}\right)
\label{eq:tcoh}
\end{equation}
This is the time during which the particle traverses a granule, 
and approximately sets the granule lifetime. It also characterises 
the timescale over which kinetic-energy differences between modes lead to a loss 
of coherence through the accumulation of relative phases $(\omega_k t)$ 
(a detailed discussion follows in Sec.~\ref{sec:refrac}).

\subsection{Simulation} 

We solve the evolution equations (Eq.~\eqref{eq:tildeGPP}) starting from an 
initial field configuration that represents a virialised DM 
halo in the solar neighbourhood. Following~\cite{Yao:2024fie}, we model 
the halo as a superposition of plane waves, each with a different velocity, and an independent random amplitude and phase. 
In the halo frame, the velocities follow a Maxwellian distribution. In the Earth frame, which moves through the halo with velocity $\mathbf v_e$, the distribution takes the form
\begin{equation}
f(\mathbf v)\propto \exp\left(-\frac{|\mathbf v+\mathbf v_e|^2}{\bar 
v^2}\right),
\label{eq:maxwell}
\end{equation}
where we take $\mathbf v_e=\bar v\,\hat{\mathbf z}$. In the rescaled 
units of Eq.~\eqref{eq:relations}, the wavevector of each plane wave is 
$\tilde{\mathbf k}=\mathbf v/\bar v$.
The initial field can therefore be written as
\begin{equation}
\tilde \psi(\tilde{\mathbf x},0)=\sum_{\tilde{\mathbf k}} 
a_{\tilde{\mathbf k}}\, z_{\tilde{\mathbf k}}\,e^{i\tilde{\mathbf 
k}\cdot\tilde{\mathbf x}},
\qquad a_{\tilde{\mathbf k}}^2\propto e^{-|\tilde{\mathbf 
k}+\tilde{\mathbf k}_e|^2},
\label{eq:rpm}
\end{equation}
with $\tilde{\mathbf k}_e=\mathbf v_e/\bar v$. Here, $a_{\tilde{\mathbf 
k}}^2$ gives the Maxwellian weight of each mode, while 
$z_{\tilde{\mathbf k}}$ is an independent complex random variable with 
zero mean and unit variance, whose modulus follows a Rayleigh 
distribution and whose phase is uniformly distributed in $[0,2\pi)$. The 
random phases cause the plane waves to interfere, producing the 
characteristic granular density pattern of the virialised field, while 
the random moduli introduce fluctuations in the amplitudes of the 
individual modes. We normalise the field such that the spatial average 
of $|\tilde\psi|^2$ over the box is unity, $\braket{|\tilde\psi|^2}=1$.

We evolve this initial configuration in a cubic box of side 
$4\lambda_{\rm dB}$ with periodic boundary conditions. The box 
represents a small, approximately homogeneous patch of the galactic 
halo. Its size is much smaller than the Jeans length, the scale above 
which gravity overcomes the pressure associated with the velocity 
dispersion. Thus, gravitational collapse does not occur on the scales 
considered here. The numerical method is described in 
Appendix~\ref{app:numerics}. 

For the $\beta=0$ case, the evolution is 
known analytically, and no numerical simulation is needed. For fixed 
$\beta$, the self-interaction part of the evolution is independent of 
$m_\phi$, and the mass enters only when the solution is converted to 
dimensional quantities using Eq.~\eqref{eq:relations}. Gravity retains a 
dependence on the mass through $\alpha\propto m_\phi^{-2}$, but is 
subdominant on the scales of the box. The changes in the granular 
structure are therefore primarily driven by the self-interactions.

From the simulations, we obtain three quantities that enter our analysis. The first is the density distribution, $R(\rho)$, which describes the occurrence of different field densities within the box. The other two follow from the autocorrelation of the density contrast:
$\delta(\mathbf x,t)\equiv\rho(\mathbf x,t)/\bar{\rho}_\phi -1$,
\begin{equation}
C(\mathbf r,\tau)=\frac{\langle\delta(\mathbf x,t)\,\delta(\mathbf x+\mathbf r,t+\tau)\rangle}
{\langle\delta^2 (\mathbf{x,t})\rangle},
\label{eq:autocorr}
\end{equation}
where $\langle\cdots\rangle$ denotes the average over positions
$\mathbf x$ in the box and over reference times $t$, so that $C(0,0)=1$.
The granule size $l_g$ is twice the separation at which the spatial autocorrelation, $C(r,0)$ averaged over the directions of $\mathbf r$, falls to $1/2$. The coherence time $\tau_g$ is the lag at which the temporal autocorrelation at a fixed point, $C(0,\tau)$, falls to $1/2$.
Thus $l_g$ is the typical size of a granule, and $\tau_g$ is the time over 
which the density at a given point stays correlated with itself. Notice that $\tau_g \sim t_{\rm coh}$ and $l_g \sim \lambda_{\rm dB}$ introduced before. These quantities characterise the density variations encountered by neutrinos and the spatial and temporal scales over which the interference pattern remains correlated. 

We use the field configuration obtained after relaxation as a proxy for the subsequent evolution of the local halo field, rather than simulating its full cosmological history. In particular, we do not model the global halo structure or the formation and dynamics of the galactic core. We identify the relaxed state with the configuration reached when the statistical properties of the field become approximately stationary.

\begin{figure}[!t]
 \centering
 \includegraphics[width=1. \linewidth]{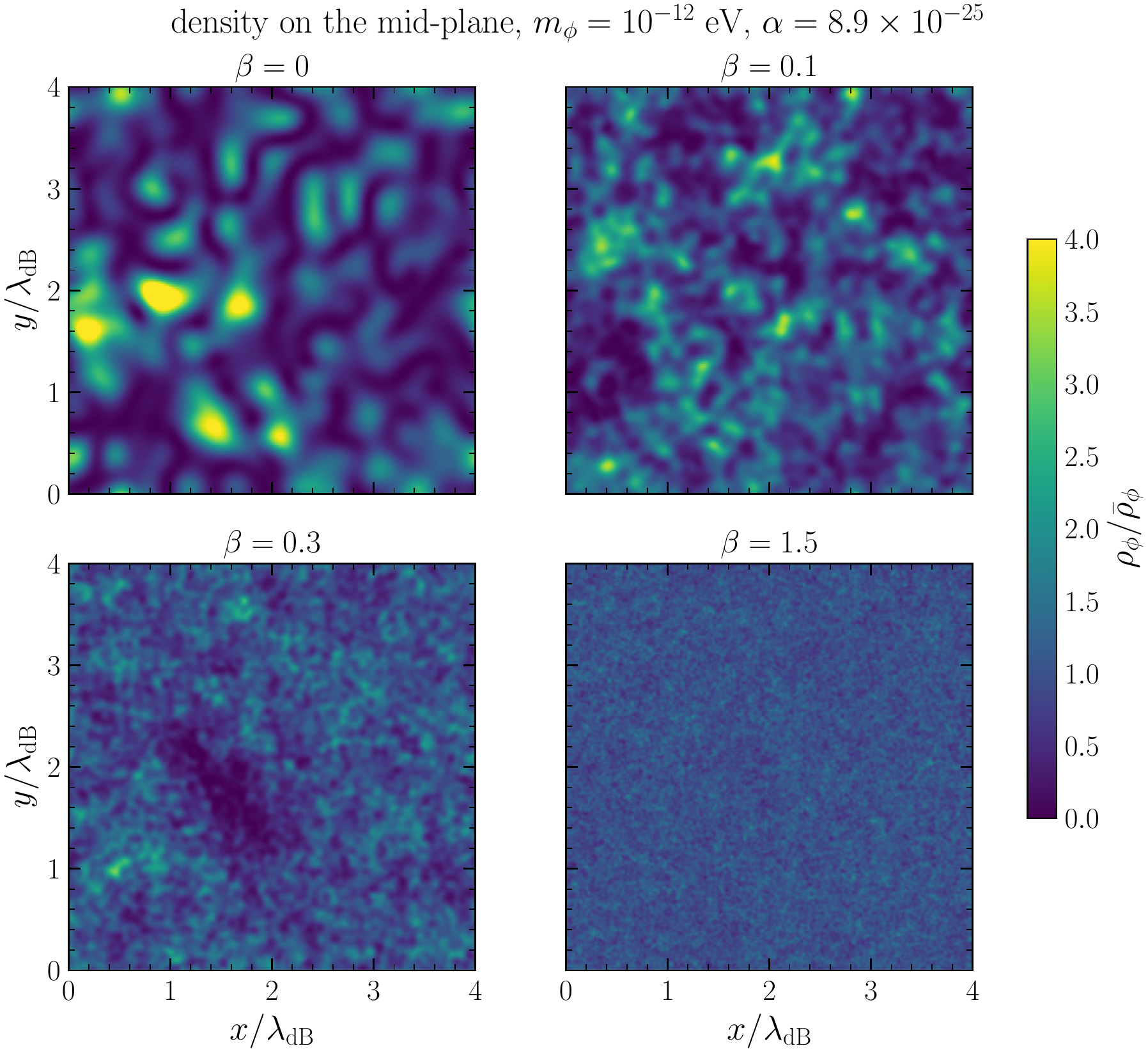}
\caption{Spatial distribution (2D slice) of energy density of the ULDM field post-evolution, for different values of self-interaction strength $\beta$. We take $m_\phi = 10^{-12}$ eV which corresponds to $\alpha = 8.9 \times 10^{-25}$.}
\label{patternb}
\end{figure}

\begin{figure}[!t]
 \centering
 \includegraphics[width=0.6\linewidth]{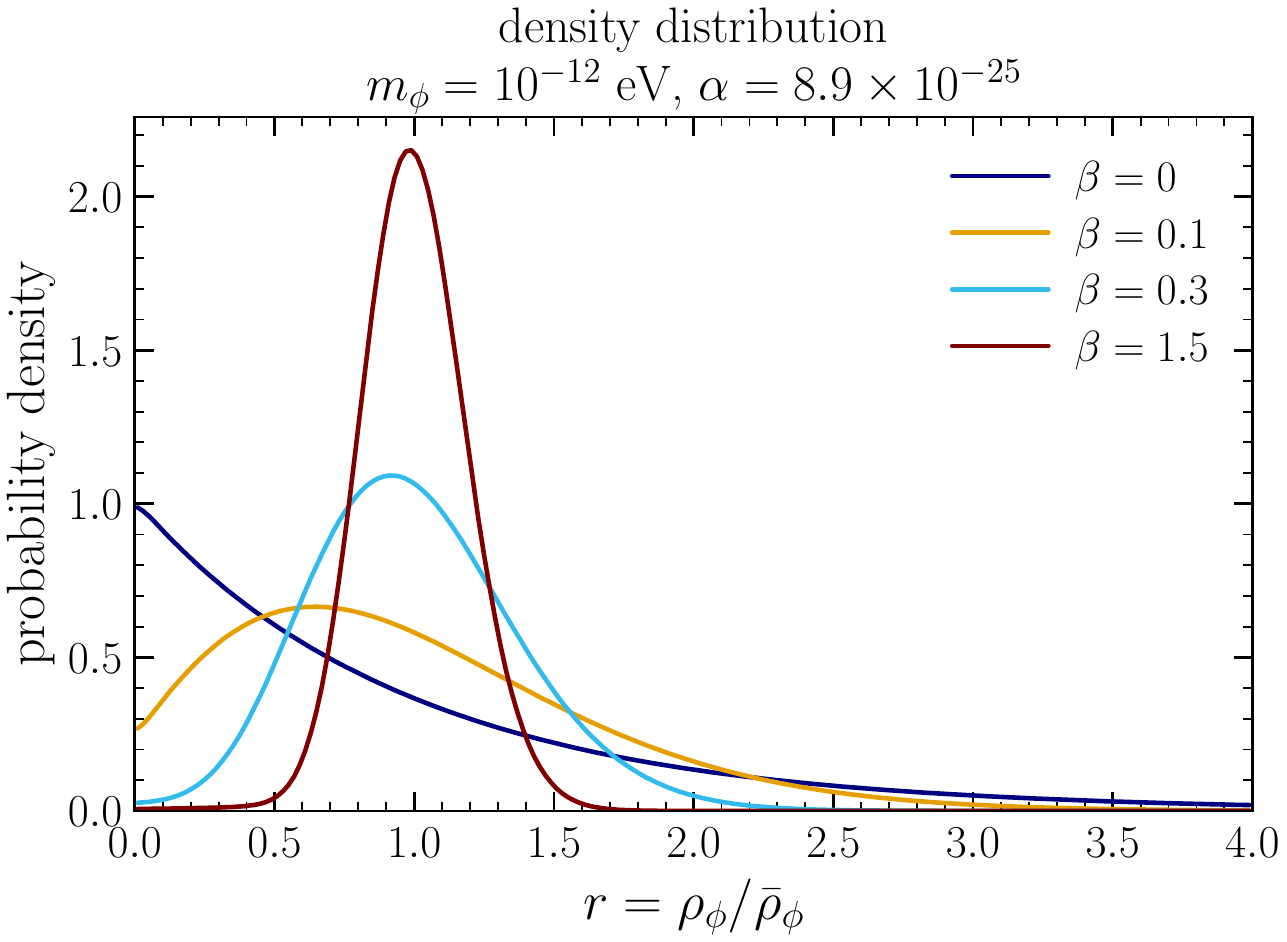}
\caption{Density distribution sampled by the detector over the observation time, for $m_\phi = 10^{-12} \, {\rm eV}$ and different values of $\beta$.}
\label{distribb}
\end{figure}

The density on a slice of the box is shown in Fig.~\ref{patternb}, while 
the corresponding distribution $R(\rho)$ is shown in 
Fig.~\ref{distribb}. We find that as $\beta$ increases, the characteristic size of 
the structures decreases and the density distribution becomes less 
contrasted. In particular, the distribution narrows and its maximum shifts towards the mean density. These changes in the field 
configuration will have crucial implications for refractive masses, as will be discussed in the following sections.

\begin{figure}[!t]
\includegraphics[width=0.49\linewidth]{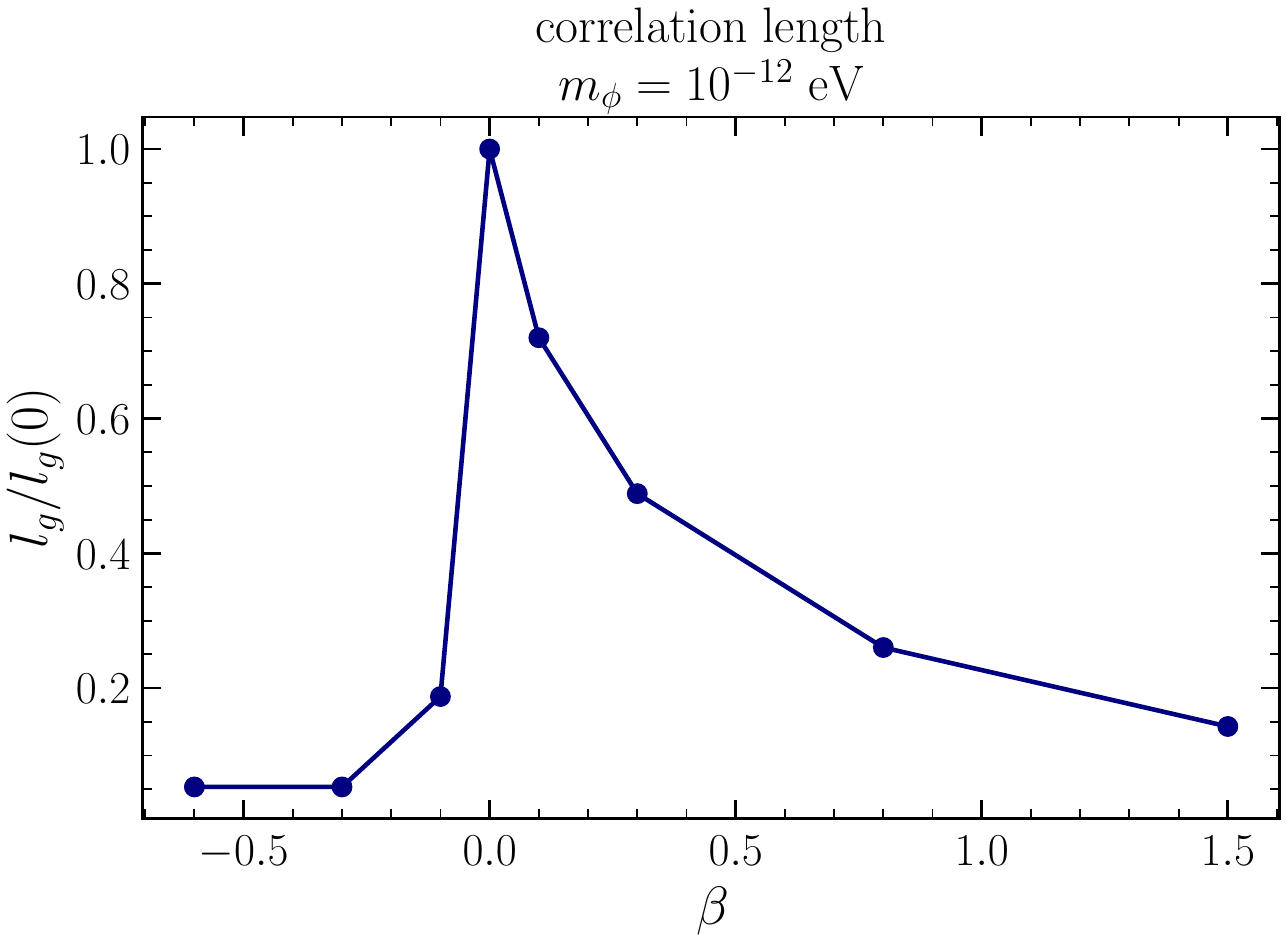}
\includegraphics[width=0.49 \linewidth]{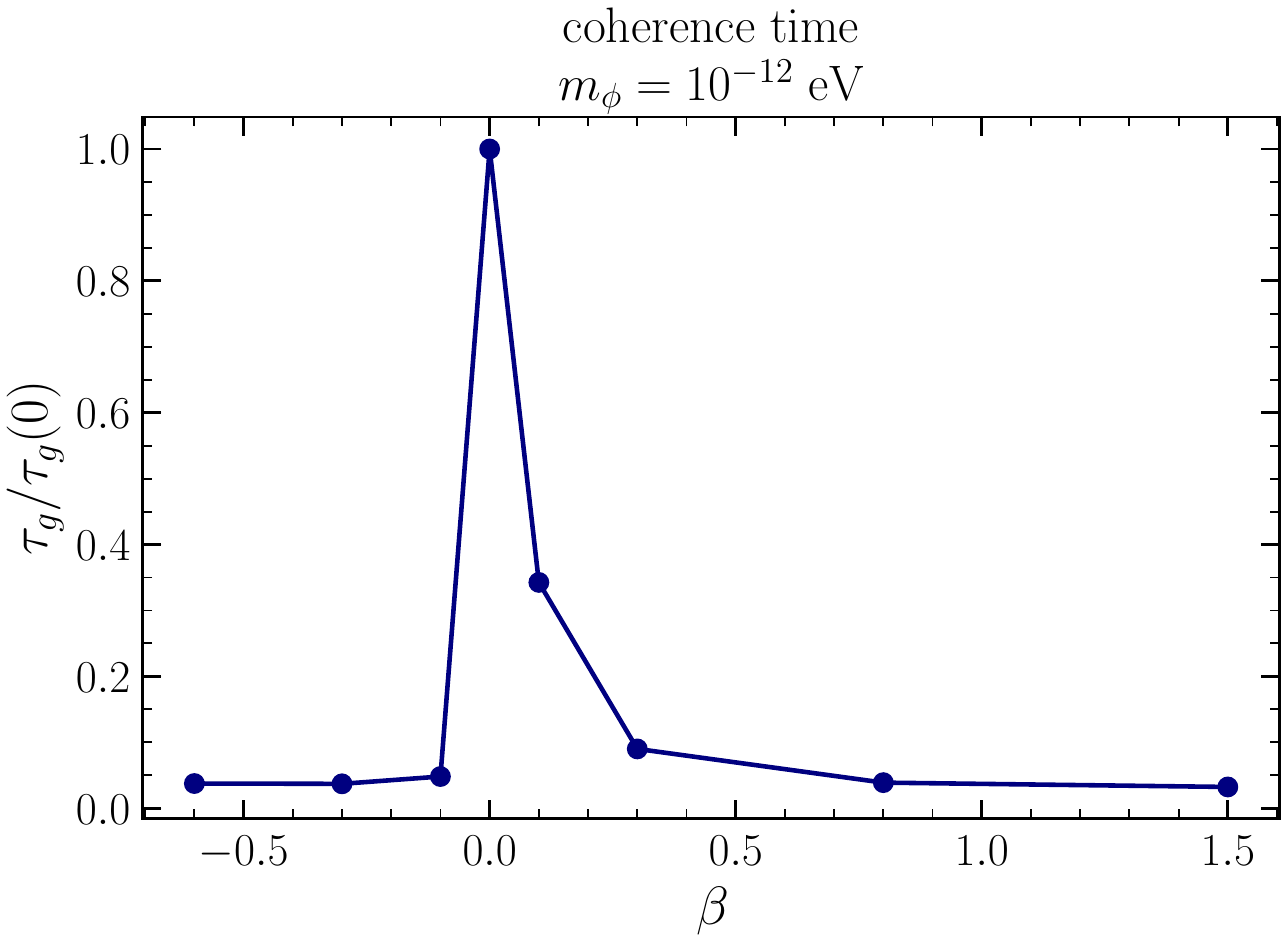}
\caption{Dependence of the granulation scale (left) and
the coherence time (right) on $\beta$.}
\label{granscale}
\end{figure}

Fig.~\ref{granscale} shows the dependence of the granule size and coherence time on $\beta$. In the absence of self-interactions ($\beta=0$), we obtain $l_g=0.356\,\lambda_{\rm dB}$ and $\tau_g=0.152\,t_{\rm coh}$. Both scales decrease with increasing $|\beta|$, irrespective of the sign of the interaction, although the coherence time exhibits a substantially stronger suppression. For $\beta=1$ and $1.5$, the granule size decreases to $20\%$ and $15\%$ of its value at $\beta=0$, respectively, while the corresponding coherence times decrease to $5\%$ and $4\%$. Following the scalings defined earlier in Eq.~\eqref{eq:relations}, $l_g/\lambda_{\rm dB}$ and $\tau_g/t_{\rm coh}$ depend only on $\beta$, and are independent of $m_\phi$. We parameterise the two scales as
\[l_g=\xi(\beta)\,\lambda_{\rm dB},\qquad \tau_g=\kappa(\beta)\,t_{\rm coh},\]
where $\xi(0)=0.356$ and $\kappa(0)=0.152$.

Although the reduction in granule size and coherence time is independent of the interaction sign, the effect on the field configuration is quite different. Repulsion suppresses density inhomogeneities, whereas attraction promotes the concentration of the field into dense, small-scale structures. We discuss these two cases separately below.

\textbf{Repulsion -- } A repulsive self-interaction acts as an effective pressure that opposes density inhomogeneities. Its effect depends on the length scale of the perturbations: at sufficiently small scales, the kinetic term dominates, and the field behaves approximately like free waves, whereas at larger scales, the repulsive interaction dominates and density perturbations propagate as sound waves. As the field relaxes, part of its mass condenses into the lowest-energy state, which has no granular fluctuations. In a self-gravitating halo, this state corresponds to the solitonic core~\cite{Indjin:2025thr}, which grows and becomes less dense with increasing coupling. However, our simulations describe a small region of this halo without a gravitational well. The lowest-energy state is therefore a single plane-wave mode that fills the entire box, and the condensed mass forms a spatially uniform background rather than a central core. The remaining density fluctuations propagate as sound waves on this background, with their characteristic scale decreasing as the repulsive coupling increases.

\textbf{Attraction -- } Attractive self-interactions promote the concentration of the field into increasingly dense, small-scale structures. The energy released during this concentration heats the remaining field, reducing both the granule size and coherence time. These reductions are substantially stronger than those produced by 
repulsive interactions. Simulations of fuzzy DM with attractive 
self-interactions~\cite{Painter:2024rnc} likewise show that, above a 
critical mass, the soliton collapses into a compact state confined to the centre of the halo. In our simulations, the field concentration is limited only by the numerical grid, and the resulting configurations are therefore not numerically converged. We use them only to illustrate the qualitative effects of attractive self-interactions (see Sec.~\ref{sec:refrac}).

\subsection{Modelling the Field}

Results of simulations can be parameterised as the sum of plane waves with momenta $k$
\begin{equation}
\phi = \frac{e^{-im_\phi t}}{\sqrt{2}m_\phi}
\sum_k A_k \exp[-i(\omega_k t - kx + \phi_k)]. 
\label{eq:field}
\end{equation}
Although gravity does not appear explicitly in this expression, its effects are encoded implicitly in the virial velocity distribution through the wavevector $k$. 
The interference among the constituent waves in Eq.~\eqref{eq:field} reproduces the pattern with granules and voids. Non-minimal effects on the field can, in turn, 
modify this pattern by changing the distributions of $A_k$ and $\phi_k$. 
This presentation gives certain insight into the physics involved. 
For $\beta \neq 0$, the self-interaction modifies the dispersion relation 
and correlates the amplitude and phases of different modes.

The terms in the exponent of Eq.~\eqref{eq:field} have the following interpretation:
\begin{itemize}
\item The phase $(kx)$ determines the spatial structure of the wave superposition. 
Summation over momenta due to the $kx$ term in the exponent would produce a single 
wave packet of the size $1/\langle k \rangle = 1/(m_\phi v)$ situated at $x = 0$.

\item The frequency $\omega_k = k^2/2m_\phi$ is the non-relativistic kinetic energy of a mode with momentum $k$, and the term $\omega_k t$ leads to the spread of the wave packet with typical time $t_{\rm coh}$. 

\item The phases $\phi_k$ are random phases with a certain distribution and encode the relative phases of the constituent modes. Their distribution determines the interference pattern, producing a granular configuration in which wave packets spread, overlap, and merge. 
\end{itemize}

Two characteristic timescales of the field govern its dynamics and interference pattern. 
\begin{itemize}
\item The coherence time introduced in Eq.~\eqref{eq:tcoh}. 

\item The Compton time period of the main mode of oscillations:
\begin{equation}
t_\phi = \frac{2\pi}{m_\phi} =
4.1 \times 10^{-5}~{\rm s} \left(\frac{10^{-10} {\rm eV}}{m_\phi}\right).
\label{eq:tcomp}
\end{equation}
\end{itemize}
The Compton length is given by $\lambda_C \equiv c t_\phi = 13 ~{\rm km}~ (10^{-10} {\rm eV}/m_\phi)$.
During one coherence time, the number of Compton oscillation periods equals 
\[t_{\rm coh} /t_\phi = 1/(\bar v^2) = 1.8 \times 10^6\,. \]
This hierarchy of timescales allows us to disentangle effects of the rapid oscillations of the field, governed by the timescale $m_\phi^{-1}$, from the much slower evolution of the interference pattern, characterised by $t_{\rm coh}$. 
Both timescales are relevant for a real scalar field (as we will see later). By contrast, for a complex scalar field, the oscillations of density with frequency $m_\phi$ are absent; the density evolves on the coherence timescale $t_{\rm coh}$.

\section{Effects of interference pattern on refractive mass and neutrino oscillations}
\label{sec:refrac}

\subsection{Oscillation probabilities and experiments}

\begin{figure}
 \centering
 \includegraphics[width=\linewidth]{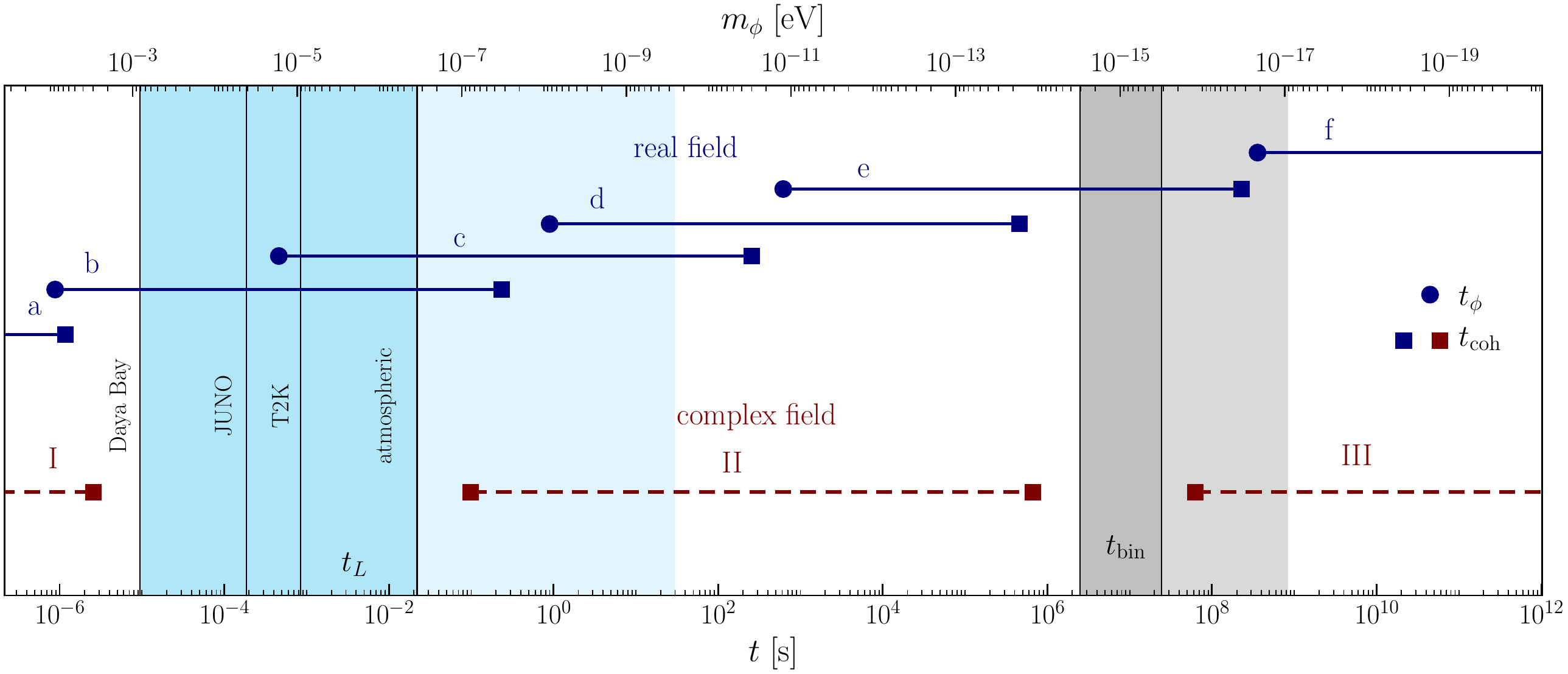}
\caption{Relevant time scales and scalar mass scales in the problem. The dark blue band shows the range of baseline times $t_{\rm L}$ of the experiments considered, from Daya Bay ($L\simeq1.7$~km) to atmospheric neutrinos ($L\simeq10^4$~km); with vertical lines marking values corresponding to Daya Bay, JUNO, T2K and atmospheric baselines. The light blue band extends the upper edge by $c/\bar v\simeq10^3$, corresponding to the spatial condition $\lambda_{\rm dB}=L$. The dark grey band shows the range of bin times $t_{\rm bin}$; the light grey band extends it to the total exposure $t_{\rm exp}$ and to the period $t_{ij}$ over which the same $\Delta m^2_{ij}$ has been measured. The upper axis gives $m_\phi$ from $t=t_{\rm coh}(m_\phi)$.  Horizontal lines show time intervals of different cases considered in the text, red for the complex field (I-III) and indigo for the real field (a-f). Circles and squares mark $t_\phi$ and $t_{\rm coh}$, respectively.}
\label{fig:scales}
\end{figure}

The field fluctuations lead to distortion and averaging of the oscillation probabilities. Therefore, we will consider experiments that observed the oscillatory dependence of the probabilities on energy.
The following temporal parameters characterise oscillation experiments (Fig.~\ref{fig:scales}).

\begin{itemize}
\item The baseline time, $t_L$, is the time of neutrino propagation from a source to a detector:
\[t_L = 3.3 \times 10^{-6}~{\rm s} \, \Big(\frac{L}{1~ {\rm km}}\Big).\]

\item The bin time, $t_{\rm bin}$, is the time interval during which a sufficient number of events are collected so that the energy spectrum can be reconstructed to analyse the data. 

\item The total exposure time, $t_{\rm exp}$, is the total duration of experiment. 
When $t_{\rm bin} < t_{\rm exp}$, temporal variations in the neutrino masses can be searched 
for by confronting the data from separate time bins.

\item $t_{ij}$ is the total time during which different experiments measure the same mass-squared difference $\Delta m^2_{ij}$.
\end{itemize}

In Fig.~\ref{fig:scales} we show $t_L$ for Daya Bay ($L = (1.6 - 1.8)$ km), JUNO ($L = 52.5$ km), T2K ($L = 295$ km) and atmospheric neutrino experiments ($L = 10^{2} - 10^{4}$ km). The upper horizontal axis gives values of $m_\phi$ according to relation $t = t_{\rm coh} (m_\phi)$,
or explicitly, $m_\phi = 2\pi / (t \bar{v}^2)$.

The observable effects depend on the relationships between the temporal scales of the experiment and those of the scalar field, which in turn depend on mass $m_\phi$. For the complex field, $t_{\rm coh}$ is the only relevant time characteristic of the field (refer to Eq.~\eqref{eq:complexfield}).
Since $t_L \ll t_{\rm bin} \sim t_{\rm exp}$ (for all neutrino experiments apart from astrophysical neutrinos, i.e., supernova and cosmic neutrinos), there are three possibilities depending on the value of $m_\phi$ summarised in Table~\ref{tab:regimes_complex}. They are determined by the temporal and spatial conditions, which are generally not equivalent and should be considered independently in each case. From case to case, one or the other leads to a stronger bound, whereas in some situations, they are equivalent.

\begin{enumerate}[label=\Roman*.]
\item $t_{\rm coh} \ll t_L$ and $\lambda_{\rm dB} \ll L$: the field variations occur much faster than propagation of individual neutrinos. 
Dividing by $c$, the spatial (second) condition can be rewritten as $ t_{\rm coh} \ll t_L c/\bar{v}$. 
This means that the first (temporal) condition is stronger. 
If we consider the case for Daya Bay, i.e., $t_L^{\rm min} = t_L^{\rm DayaBay}$, 
this condition is satisfied for $m_\phi > 10^{-2}$ eV. 
In this case, variations in the field should be accounted for when propagating individual neutrinos. 

\item $t_{\rm coh} \gg t_L$, $\lambda_{\rm dB} \gg L$: the field varies much slower 
than the propagation of an individual neutrino, and structures are much bigger than the baseline. 
The second condition can be rewritten as $t_{\rm coh} \gg t_L c/\bar{v}$. 
So, now the spatial condition is 3 orders of magnitude stronger, which is shown by 
the light blue band in Fig.~\ref{fig:scales}.

Taking $t_L = t_L^{\rm max} = t_L^{\rm atm}$, we find that this condition is satisfied 
for all experiments for $m_\phi < 10^{-10}$ eV. The temporal condition gives a weaker bound: $m_\phi < 10^{-7}$ eV.
This regime holds for $t_{\rm coh} < t_{\rm bin}$. In this range, the evolution of individual neutrinos is not affected, 
and the oscillation probability should be averaged at the ensemble level.

\item $t_{\rm coh} \gg t_{\rm bin}$, $x_{\rm bin} \equiv \bar{v}t_{\rm bin} \ll \lambda_{\rm dB}$: 
time scale of the field variations is much bigger than the duration of a given experiment, and spatial variations have bigger scales than the distance the experimental 
set up crosses during the experiment. Here, both conditions are equivalent: indeed, multiplying the first condition by $\bar{v}$ yields the second. For $t_{\rm bin} = 1$ year the condition is satisfied at $m_\phi < 10^{-16}$ eV. With increasing $\beta$, the size of the granules decreases, and therefore the upper edge of this region moves to a smaller $m_\phi$. 
On the other hand, one can strengthen the limit by analysing 
different experiments that measure the same mass-squared difference in different periods of time. E.g., $\Delta m^2_{32}$ was measured from 1990 to the present, i.e., during about 36 years. $\Delta m^2_{21}$ was measured from 2002 (KamLAND) to the present (JUNO), i.e., during 24 years. Therefore, the fact that $\Delta m^2_{ij}$ does not change during these periods allows us to improve the upper bounds by 1.5 orders of magnitude: $m_\phi < 3 \times 10^{-18}$ eV. 
\end{enumerate}

There are two transition regions around two critical values of $m_\phi$ determined by the conditions 
$t_{\rm coh} = t_L$ and $t_{\rm coh} = t_{\rm bin}$.

\begin{table}[t]
\caption{Regimes of field variations for a complex scalar field, in order of
decreasing $m_\phi$. The interference pattern evolves on the timescale
$t_{\rm coh}$. The notations used are as follows. \textit{path}: averaged along each neutrino trajectory.
\textit{ensemble}: averaged over the neutrinos detected during $t_{\rm bin}$.
\textit{frozen}: constant during $t_{\rm bin}$. The boundaries follow from
$t_{\rm coh}=t_L$ and $t_{\rm coh}=t_{\rm bin}$. With self-interactions, the boundaries set by $t_{\rm coh}$ shift to smaller $m_\phi$ by a factor of $
\kappa(\beta)$.}
\label{tab:regimes_complex}
\begin{ruledtabular}
\begin{tabular}{cllc}
Regime & Timescale condition & Interference pattern & $m_\phi$ (eV) \\
\hline
I   & $t_{\rm coh}\ll t_L\ll t_{\rm bin}$  & path                         & $\gtrsim 10^{-3}\,(1\,{\rm km}/L)$ \\
II  & $t_L\ll t_{\rm coh}\ll t_{\rm bin}$  & ensemble                  & $10^{-16}\,(1\,{\rm yr}/t_{\rm bin})$--$10^{-3}\,(1\,{\rm km}/L)$ \\
III & $t_L\ll t_{\rm bin}\ll t_{\rm coh}$  & frozen                       & $ \lesssim 10^{-16}\,(1\,{\rm yr}/t_{\rm bin}) $ \\
\end{tabular}
\end{ruledtabular}
\end{table}

In general, computations of oscillation probabilities in the fluctuating field proceed in two steps.

\begin{enumerate}
  \item Computation of the probability, $P(t_0, t_L)$, for an individual neutrino produced at the moment of time $t_0$, in an experiment with baseline time $t_L$. For a given experiment (with a known position in space), $t_0$ completely determines the configuration of the field that the neutrino crosses.
  \item Averaging of the probability over an ensemble of neutrino events detected during the observation time bin $t_{\rm bin}$:
\begin{equation}
\langle P(t_0, t_L) \rangle = \frac{1}{t_{\rm bin}}
\int_{t_{\rm in}}^{t_{\rm in} + t_{\rm bin}} dt_0 P(t_0, t_L),
\label{eq:ensembleav}
\end{equation}
where $t_{\rm in}$ is the time of beginning of observations. 
This formula is valid if the characteristics of the experiment do not change during $t_{\rm bin}$. In general, parameters of the experiment such as neutrino flux $F_\nu(t_0)$, efficiency of detection $D(t_0)$, etc., change and therefore 
\begin{equation}
\langle P(t_0, t_L) \rangle = \frac{1}{\mathcal{N}}
\int_{t_{\rm in}}^{t_{\rm in} + t_{\rm bin}} dt_0 F_\nu(t_0) D(t_0)P(t_0, t_L),
\label{eq:ensembleavg}
\end{equation}
with normalisation factor $$\mathcal{N} = \int_{t_{\rm in}}^{t_{\rm in} + t_{\rm bin}} dt_0 F_\nu(t_0) D(t_0).$$
It is this averaged probability that is used to extract neutrino oscillation parameters. 
\end{enumerate}

In our framework~\cite{Sen:2023uga}, since all the neutrino masses are proportional to the same field squared, the mixing is independent of the field. It is determined by the couplings of the scalar to the refractive mass eigenstate $\nu_i$, and it arises from a mismatch between interactions of neutrinos with W bosons and with the scalar field.
In the case of a complex field, according to Eq.~\eqref{eq:refaction}, the difference of refractive mass squared can be written as
\begin{equation}
\Delta m^2_{\rm ref}(t) = \frac{\Delta g_\phi^2}{2m_\phi^2}
|\psi (t)|^2 = \frac{\Delta g_\phi^2}{2m_\phi^2} \rho(t),
\label{eq:dms-com}
\end{equation}
where $\Delta g_\phi^2 = g_{\phi 2}^2 - g_{\phi 1}^2$. Since mixing does not depend on the scalar field~\cite{Sen:2023uga},
the integration of the flavour evolution equation for the individual
neutrino in the basis $(\nu_2, \nu_1)$ is trivial, leading to the standard form 
of the oscillation probabilities, e.g. the survival one 
\begin{equation}
P(t_0, t_L) = 1 - \sin^2 2 \theta \sin^2 (\phi_{\rm osc}(t_0, t_L)/2)
\label{eq:prob-osc}
\end{equation}
with the oscillation phase
\begin{equation}
\phi_{\rm osc} = \int_{t_0}^{t_0 + t_L}dt\frac{\Delta m^2_{\rm ref}(t)}{2E} = \frac{\Delta g_\phi^2}{4m_\phi^2 E} \int_{t_0}^{t_0 + t_L} dt \rho(t). 
\label{eq:phase-osc}
\end{equation}

Let us present the averaged oscillation probabilities for the three regimes, as described in Table~\ref{tab:regimes_complex}.

\begin{enumerate}[label=\Roman*.]
  \item If $t_L \gg t_{\rm coh}$ (long baseline), the integral for the phase Eq.~\eqref{eq:phase-osc} should be computed explicitly
\begin{equation}
\phi_{\rm osc} = \frac{t_L}{2E} \frac{1}{t_L}\int_{t_0}^{t_0 + t_L} dt \Delta m^2_{\rm ref}(t) =
\overline{\Delta m^2_{\rm ref}}(t_0)\frac{t_L}{2E},
\label{eq:phase-osc2}
\end{equation}
where $\overline{\Delta m^2_{\rm ref}}$ is the mass-squared difference averaged over a given neutrino trajectory. For long baselines (so that many granules are crossed), $\overline{{\Delta m^2_{\rm ref}}}(t_0)$ depends on $t_0$ very weakly. Essentially, $\overline{\Delta m^2_{\rm ref}}$ is constant, as in the standard case of vacuum masses. Therefore, averaging over the ensemble does not change the result. The equality Eq.~\eqref{eq:phase-osc} can be rewritten as
\begin{equation}
\phi_{\rm osc} = \frac{\Delta g_\phi^2 t_L}{4m_\phi^2 E}~
\frac{1}{t_L}\int_{t_0}^{t_0 + t_L}  dt \,\rho(t) =\frac{\Delta g_\phi^2 t_L}{4m_\phi^2 E} \braket{\rho(t)}. 
\label{eq:phase-osc1}
\end{equation}
That is, the average energy density along the baseline determines the phase. Then averaged probability is given in Eq.~\eqref{eq:prob-osc} with $\phi_{\rm osc}$ from Eq.~\eqref{eq:phase-osc1}. Distortion of the probability is not expected.
In practice, the condition $t_L \gg t_{\rm coh}$ can be satisfied only for very large masses and very-long-baseline experiments. Indeed, according to Eq.~\eqref{eq:tcoh}
\begin{equation}
\frac{t_L}{t_{\rm coh}} = 4.3 \times 10^{-8}
\left(\frac{L}{{\rm km}} \right)
\left(\frac{m_\phi}{10^{-10}{\rm eV}} \right).
\label{eq:timeratio}
\end{equation}
For the atmospheric neutrinos with $L \sim 10^4$ km, the inequality 
$t_L/t_{\rm coh} \gg 1$ is satisfied for 
$m_\phi > 10^{-6}$ eV.
For reactor neutrino experiments, the mass required to
avoid a distortion is $m_\phi > 10^{-2}$ eV, which is beyond the range for refractive mass. Using Eq.~\eqref{eq:tcoh} and setting $t_{\rm coh}=t_{\rm L}$, with $v_e = 220~{\rm km/s}$, we get critical mass as- 
\begin{equation}
m^L_\phi = \frac{2\pi}{t_L\bar v^2}=2.3\times10^{-3}\, ~{\rm eV}\left(\frac{1~{\rm km}}{L}\right).
\label{eq:mL}
\end{equation}

\item If $t_L \ll t_{\rm coh}$ (short baseline), the density, and consequently 
$\Delta m_{\rm ref}^2$, is practically a constant during neutrino propagation, so that according to Eq.~\eqref{eq:phase-osc}
\begin{equation}
\phi_{\rm osc}(t_0, t_L) = \frac{\Delta m^2_{\rm ref}(t_0) t_L}{2E} \approx \frac{\Delta m^2_{\rm ref}(t_0) L}{2E}.
\label{eq:phase-oscll}
\end{equation}
The integration over time in Eq.~\eqref{eq:ensembleav} can be reduced to an integration over the density, using the density distribution that neutrinos encounter during the experiment. Indeed, an experimental setup crosses many granules, and therefore many regions $j$ with the same density $\rho$. Total time the setup spends in the regions with density $[\rho ,\,\rho + d\rho]$ equals $$ \Delta t_\rho = \sum_j \left|\frac{dt}{d\rho} \right| d\rho.$$ Then the integral over time can be written as 
\begin{equation}
\langle P(\rho) \rangle = \int_0^\infty d\rho P(\rho) \frac{1}{t_{\rm bin}}\sum_{j(\rho)} \left|\frac{dt}{d\rho}\right|_{j(\rho)}, 
\label{eq:int-rho}
\end{equation}
where the product of factors
\begin{equation}
R(\rho) \equiv \frac{1}{t_{\rm bin}} \sum_{j(\rho)} \left| \frac{dt}{d\rho}
\right|_{j(\rho)}
\label{eq:distr}
\end{equation}
gives the probability that a setup encounters a region with density $\rho$ during the bin time. So, $R(\rho)$ is the normalised density distribution along the trajectory of the setup. Now Eq.~\eqref{eq:int-rho} can be rewritten as
\begin{equation}
\langle P(\rho) \rangle = \int_0^\infty d\rho R(\rho) P(\rho).
\label{eq:int-rho2}
\end{equation}

Thus, the effect of the interference pattern on oscillations 
is encoded in the distribution $R(\rho)$. The distribution has a maximum at characteristic density, $\rho_{\rm max}$, so that $\Delta m_{\rm ref}^2(\rho_{\rm max})$ can be regarded as an effective vacuum mass-squared splitting.

The width of the distribution determines the extent of the averaging effect: the narrower the distribution, the weaker the averaging. The distribution $R(\rho)$ and the spatial scale of the structures are the central objects of our study. Since $\Delta m_{\rm ref}^2 \propto \rho$ 
averaging over $\rho$, Eq.~\eqref{eq:int-rho2} can be represented as averaging over $\Delta m_{\rm ref}^2$: 
\begin{equation}
\langle P(\Delta m_{\rm ref}^2) \rangle = \int_0^\infty d(\Delta m_{\rm ref}^2) R(\Delta m_{\rm ref}^2) P(\Delta m_{\rm ref}^2)
\label{eq:int-rho3}
\end{equation}
with normalised distribution $R(\Delta m_{\rm ref}^2) = R(\rho) d\rho/ d(\Delta m_{\rm ref}^2)$
In turn, since the oscillation phase equals 
$(\Delta m_{\rm ref}^2 x/4E)$, the averaging over 
$\Delta m_{\rm ref}^2$
is equivalent to usual averaging over baselines $x$, as discussed in \cite{Cheek:2025kks}.
Recall that the description Eq.~\eqref{eq:int-rho3} is valid if 
during $t_{\rm bin}$, neutrinos or experimental setups cross many granules, corresponding to $t_{\rm bin} \gg t_{\rm coh}$.

\item For $t_{\rm coh} \gg t_{\rm bin}$, the density is frozen during the exposure. Therefore, the only effect is a rescaling of $\Delta m^2$. In this connection, let us define the critical scalar mass 
$m_\phi^{\rm cr}$ as the mass for which the duration of the experiment, $t_{\rm bin}$, equals the coherence time: 
\begin{equation}
t_{\rm bin} = t_{\rm coh}. 
\label{eq:tt}
\end{equation}
Equivalently, this condition can be expressed by requiring that the distance travelled by the experimental setup during
$t_{\rm bin}$, denoted by $l_{\rm bin}$, equals the de Broglie wavelength, which characterises the size of a granule: 
\begin{equation}
l_{\rm bin} = \lambda_{\rm dB}. 
\label{eq:xx}
\end{equation}
The two conditions are equivalent if the velocity of the solar system in the galaxy equals the virial velocity $v_S = v_e = \bar{v}$. 
Using Eqs.~\eqref{eq:tcoh} and Eq.~\eqref{eq:tt}, we obtain
\begin{equation}
 m_\phi^{\rm cr} = \frac{2\pi}{t_{\rm bin}\bar{v}^2} = 2.4 \times 10^{-16}{\rm eV}\left(\frac{1~{\rm yr}}{t_{\rm bin}}\right).
\label{eq:mcrit}
\end{equation}
Here we have taken $v_e = 220$ {km/s}. 
The number of coherence time periods in the exposure is 
$N = t_{\rm bin}/t_{\rm coh} = m_\phi/m_\phi^{\rm cr}$.
For $m_\phi \ll m_\phi^{\rm cr}$, or $t_{\rm bin} \ll t_{\rm coh}$ the 
effect of time variations is absent. Conversely, 
for $m_\phi \gg m_\phi^{\rm cr}$, we have $t_{\rm bin} \gg t_{\rm coh}$, 
and the oscillation probabilities are distorted or completely averaged. 
Our simulations show that the transition between these two regimes spans approximately one order of magnitude in $m_\phi$. 
Thus, values of $m_\phi < m_\phi^{\rm cr}$ are allowed. 
The critical value of the mass Eq.~\eqref{eq:mcrit} depends on the duration of the experiment 
and also on the accuracy with which an experiment measures 
the oscillation probability, shifting $m_\phi^{\rm cr}$ to smaller values as accuracy increases. 

If the experimental setup crosses only one or a few granules, 
the result depends on the specific density distribution, 
and no general formula can be obtained. 
In the estimates above, we assumed that the de Broglie wavelength 
and the coherence time set the spacetime scales of structures. This is true if self-interactions can be treated as a perturbation that does not strongly affect scales. 

In the preceding estimates, we used $t_{\rm coh}$ as the characteristic timescale of the interference pattern. However, the relevant timescale for the evolution of the density at a fixed spatial point is its correlation time, $\tau_g=\kappa(\beta)\,t_{\rm coh}$. This timescale is already shorter than $t_{\rm coh}$ for $\beta=0$ and decreases further with increasing $|\beta|$. The conditions $t_L=\tau_g$ and $t_{\rm bin}=\tau_g$ are therefore reached at
\begin{align}
    m^L_\phi &= \kappa(\beta)\frac{2\pi}{t_L\bar v^2}=2.3\times10^{-3}\,\kappa(\beta)~{\rm eV}\left(\frac{1~{\rm km}}{L}\right),\\
    m_\phi^{\rm cr} &= \kappa(\beta)\frac{2\pi}{t_{\rm bin}\bar v^2}=2.4\times10^{-16}\,\kappa(\beta)~{\rm eV}\left(\frac{1~{\rm yr}}{t_{\rm bin}}\right).
\end{align}
Consequently, all mass boundaries determined by the evolution of the interference pattern, including those shown in Fig.~\ref{fig:scales} and Tables~\ref{tab:regimes_complex} and~\ref{tab:regimes_real}, shift to lower masses by a factor of $\kappa(\beta)$.
\end{enumerate}

\subsection{Results for oscillation probabilities}

Using the field configurations obtained from our simulations, 
we determined the density as a function of distance along neutrino trajectories. 
Using a large number of trajectories, we obtained
the probability that neutrinos encounter a given density of DM during 
long enough observation time. According to Eq.~\eqref{eq:dms-com}, 
this yields the distribution $R(\Delta m_{\rm ref}^2) \propto R(\rho)$. Substituting this distribution into Eq.~\eqref{eq:int-rho2}, we computed the averaged oscillation probabilities. A detailed fit to experimental data using these probabilities 
is beyond the scope of this paper. We therefore draw only semi-quantitative conclusions based on the figures.

\begin{figure}
 \centering
 \includegraphics[width=1.0 \linewidth]{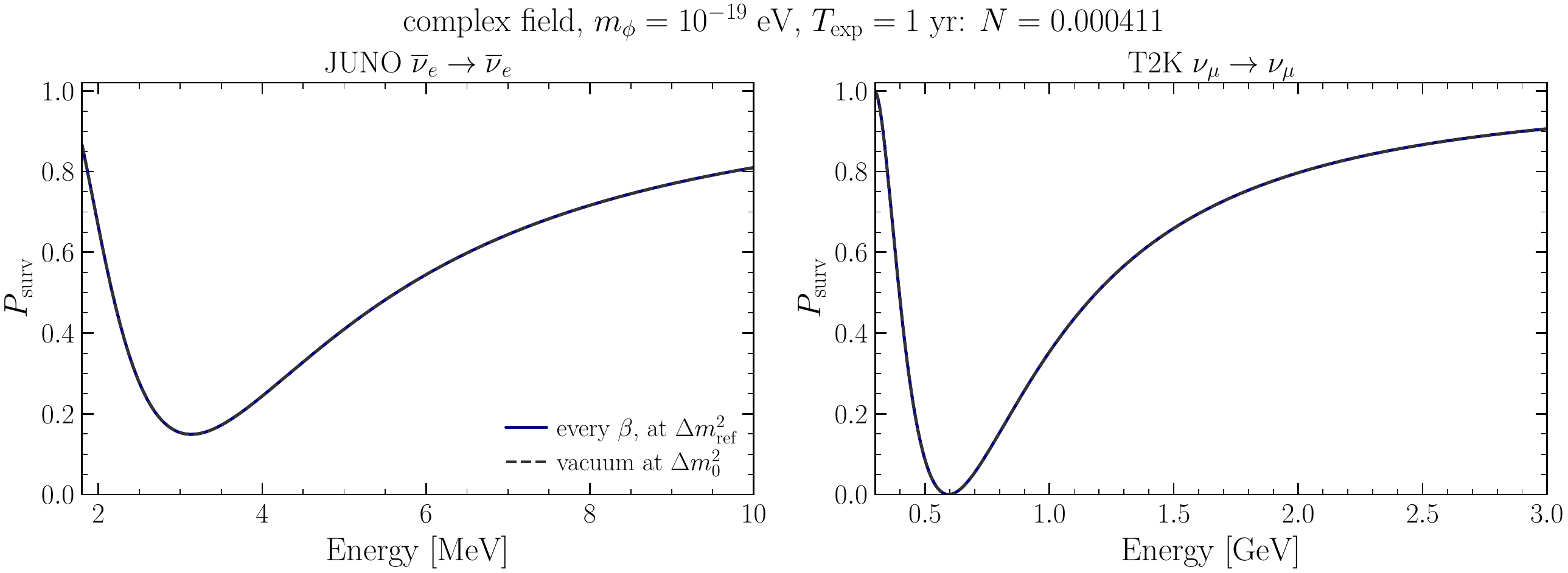}
\caption{Oscillation survival probability as a function of neutrino energy 
for $m_\phi = 10^{-19}$ eV and different values of $\beta$. Left panel for JUNO, right panel for T2K.}
\label{prob19}
\end{figure}

\begin{figure}
 \centering
 \includegraphics[width=1.0 \linewidth]{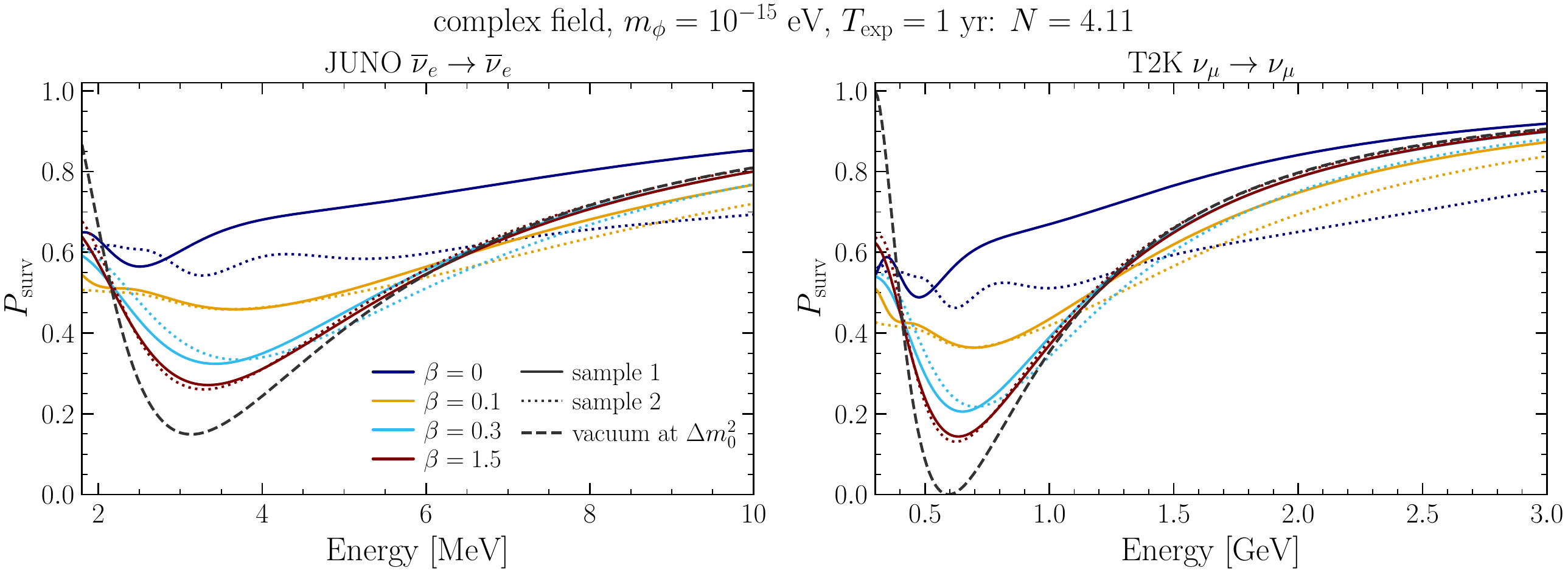}
\caption{Oscillation survival probability as a function of neutrino energy for 
$m_\phi = 10^{-15}$ eV, different values of $\beta$.
For $m_\phi = 10^{-15}~\mathrm{eV}$, the number of independent field granules sampled during the observation period is only $\mathcal{O}(1)$; therefore, the oscillation probabilities are shown for two representative realisations of the field. Left panel for JUNO, and right panel for T2K.}
\label{prob15}
\end{figure}

\begin{figure}
 \centering
 \includegraphics[width=1.0 \linewidth]{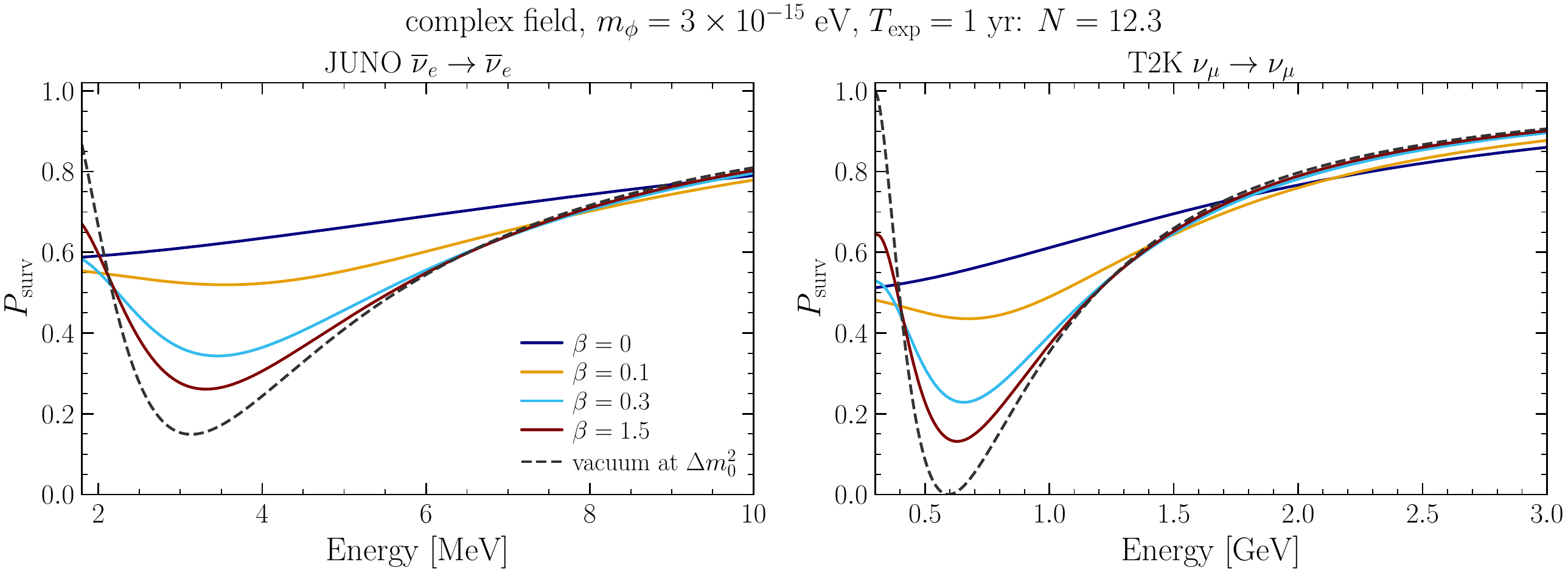}
\caption{The same as in Fig. \ref{prob19} but for $m_\phi = 3 \times 10^{-15}$ eV.}
\label{prob315}
\end{figure}

\begin{figure}
 \centering
 \includegraphics[width=1.0 \linewidth]{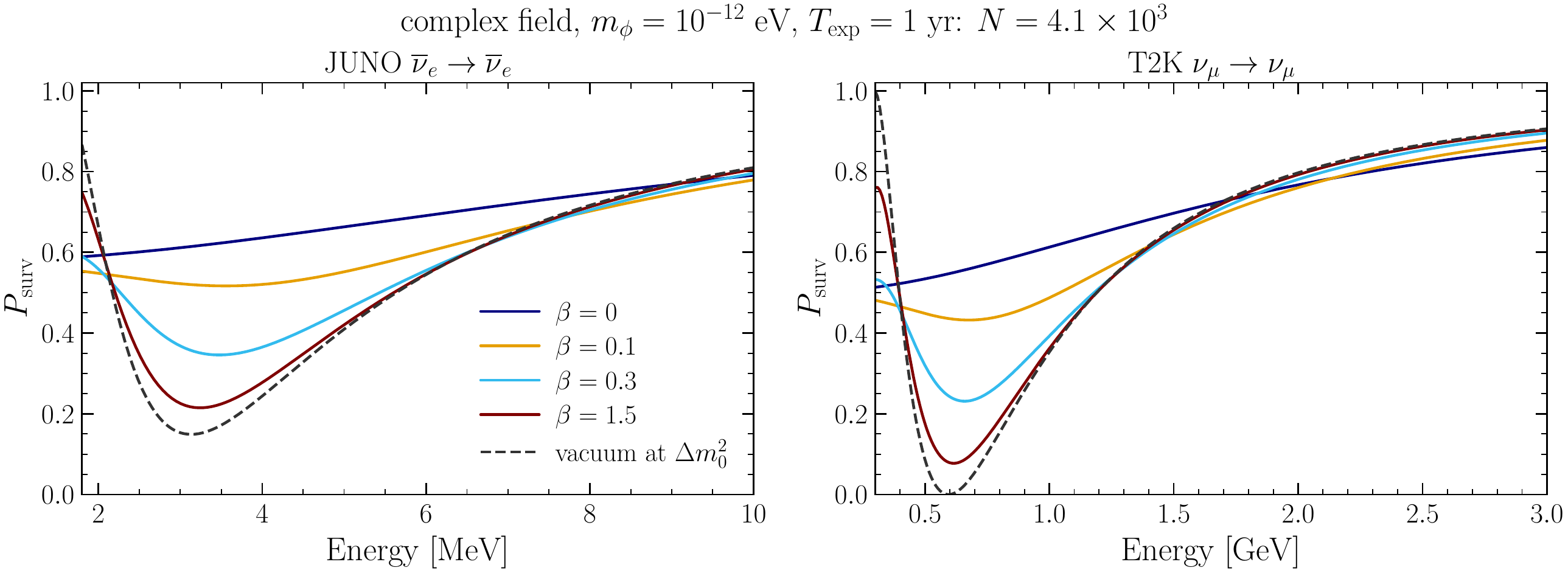}
\caption{The same as in Fig. \ref{prob19} but
for $m_\phi = 10^{-12}$ eV.}
\label{prob12}
\end{figure}

There are two effects of the field self-interactions:

\begin{enumerate}

\item Smoothing of density contrast -- Repulsive self-interactions deplete the regions of very low density, 
shifting the peak of the distribution away from zero and towards the mean density. 
This shift reduces the overall spread of the distribution. Since the relative 
spread determines the extent to which oscillations are washed out, the averaging effect weakens. 
Extrapolating to larger $\beta$ should drive the distribution towards a delta function, thereby recovering the vacuum oscillation results.

\item Decrease in the correlation length and therefore scale of the structures, as shown in Fig.~\ref{granscale}. 

\end{enumerate}

In Figs.~\ref{prob19}-\ref{prob12} we show the survival oscillation probabilities due to the refractive mass as a function of energy for different values of $m_\phi$ and self-interaction strength $\beta$. The $\bar{\nu}_e - \bar{\nu}_e$ probability is computed for the JUNO experiment~\cite{JUNO:2022mxj, JUNO:2025gmd}, while the ${\nu}_\mu - {\nu}_\mu$ probability is evaluated for the T2K experiment~\cite{T2K:2023smv} in the two-flavour approximation. We consider only repulsive self-interactions. For comparison, we also show the probabilities obtained using the vacuum mass alone. We treat $\Delta m^2$ as a free parameter and, for each case, adjust it to obtain 
the best fit to the oscillation probability in the presence of the refractive mass. This means that by adjusting the coupling $g_i$ 
we can obtain the observed $\Delta m^2$. The corresponding results for KamLAND \cite{KamLAND:2008dgz} and T2HK \cite{Hyper-Kamiokande:2018ofw}, \cite{Hyper-KamiokandeWorkingGroup:2014czz} are expected to be similar.

Fig.~\ref{prob19} corresponds to $m_\phi = 10^{-19}$ eV, which is much smaller than the critical mass even for $t_{\rm bin} = 10$ years (the case III). Consequently, no observable effect from variations is expected, and an appropriate adjustment of the couplings $g_i^2$ can reproduce the probabilities due to the vacuum masses.

Fig.~\ref{prob15} shows the oscillation probabilities for $m_\phi = 10^{-15}~\mathrm{eV}$, which equals approximately the critical mass $m_\phi^{\rm cr}$ for a one-year observation period. 
Since the number of independent field granules sampled during 
this period is $\mathcal{O}(1)$, neither the coupling adjustment nor ensemble averaging is applicable. We therefore show two randomly drawn realisations of the field to illustrate the resulting uncertainty in the oscillation probabilities. 
According to Fig.~\ref{prob15}, with an increase in $\beta$, the energy dependence of oscillation probability with refractive mass approaches that with vacuum mass. Moreover, curves for the two samples become closer to each other. The latter occurs because, with 
an increase of $\beta$, the spacetime scale of structures becomes smaller, and therefore the effective number of granules the detector crosses during the experiment increases.

Fig.~\ref{prob315} corresponds to the mass $m_\phi =3 \times  10^{-15}$ eV and about 10 crossed granules. Dependence on specific trajectories becomes weaker, and averaging becomes stronger.

For $m_\phi = 10^{-12}$ eV (Fig.~\ref{prob12}), which is much bigger than the critical mass, neutrinos cross $4.1 \times 10^{3}$ granules. 
Therefore, the averaging is very strong (and nearly complete). 
The averaging effect becomes even stronger as energy decreases, and the probability converges to $\approx 1/2$.

Results of Figs.~\ref{prob315} and \ref{prob12} are close to each other, so that increase of $m_\phi$ above $3 \times 10^{-15}$ eV shows a kind of ``saturation'': the probabilities practically do not change with $m_\phi$. Deviation from the vacuum case 
can be quantified by examining the difference in probabilities, $\Delta P$, at the first oscillation minimum. 
According to Fig.~\ref{prob12}, for $\beta = 0, 0.1, 0.3, 1.5$ the differences 
are $\Delta P = 0.48, 0.37, 0.21, 0.07$ for JUNO and $\Delta P = 0.57, 0.42, 0.23, 0.07$ for T2K. 
Extrapolating results to larger $\beta$, we find that $\Delta P = 0.05$ can be achieved at $\beta = 2.2 - 2.4$ in both experiments. Thus, elimination of bounds on the refractive mass for $m_\phi > 3 \times 10^{-15}$ eV requires $\beta > 2 - 2.5$.

As shown in Figs.~\ref{prob19}-\ref{prob12} for JUNO, at $\beta = 0$, variations 
of refractive masses lead to strong averaging of probabilities, and therefore, 
to a strong deviation from probabilities of vacuum oscillations. Effects in KamLAND are similar to JUNO, 
and therefore the interval of masses of $m_\phi = (10^{-15},10^{-9})$ eV is already excluded, 
confirming results of \cite{Cheek:2025kks} and \cite{Chathirathas:2026bgu}. We emphasise that this 
conclusion holds even in the absence of rapid temporal variations in the field.

We present here the results for the JUNO and T2K experiments, which measure probabilities with high accuracy. 
Notice that results for JUNO and T2K are rather similar even though their baselines 
differ by a factor of 5.6. In fact, in regime II, the results depend little on $L$ and $t_0$ 
and therefore can be applied to other experiments (MINOS, NOvA, KamLAND). 
In all existing experiments, the energy dependence of oscillations was observed within the first oscillation minimum (maximum). 
The difference appears at the upper border of region II, which depends on $L$ through the critical value of $m_\phi$. 
Consequently, the difference in oscillation effects 
may appear in the transition region 
$m_\phi = (10^{-10} - 10^{-3})$ eV, where experiments can be in two different regimes, 
e.g., T2K is in the pattern ($\rho$) averaged region, while Daya Bay is 
in the ensemble-averaged region. However, self-interactions may reduce this difference.

\subsection{Real scalar field}
\label{subsec:real_field}

Let us consider a real scalar field with mass and self-interaction 
described by a potential $V=\tfrac12m_\phi^2\phi^2+\tfrac14\lambda\phi^4$. 
In the non-relativistic case, the field can be represented as
\begin{equation}
\phi = \frac{1}{\sqrt{2}m_\phi}
\left( \psi \, e^{-i m_\phi t } + \psi^* \, e^{ i m_\phi t} \right), 
\label{eq:realphi}
\end{equation}
where the rapidly oscillating factors are separated from 
the slowly varying function $\psi$, 
which obeys Eqs.~\eqref{eq:GPP} with $\lambda/(2m_\phi^3) \rightarrow 3\lambda/(4m_\phi^3)$, 
i.e. $\beta=3\lambda\bar\rho/(4m_\phi^4\bar v^2)$. 

The refractive mass is determined by the field squared:
\begin{equation}
\phi^2=\frac{\rho}{m_\phi^2}\left[1+\cos(2m_\phi t+\vartheta)\right],
\label{eq:phi2real}
\end{equation}
where $\rho = |\psi|^2$ is the slowly varying density-factor and 
$\vartheta \equiv -2\operatorname{Arg}\psi$ is the phase of $\psi$. 
Thus, in addition to the interference pattern variations, the field squared oscillates 
in time with maximal amplitude and period $t_\phi/2$. According to Eq.~\eqref{eq:phi2real}, the refractive mass-squared splitting equals 
\begin{equation}
\Delta m^2_{\rm ref}(\mathbf{x},t)=\overline{\Delta m^2}(\mathbf{x},t) \left[1+\cos(2m_\phi t+\vartheta)\right],
\label{eq:modulation}
\end{equation}
where $\overline{\Delta m^2}\propto\rho(\mathbf x,t)$ 
defined analogously to Eq.~\eqref{eq:dms-com} follows the slowly varying pattern. 

In this case, the Compton scale 
$t_\phi$ (\ref{eq:tcomp}) is involved. It determines the new critical mass, $m_{\phi,\rm Compton}^L$, 
by the equality $t_\phi = 2\pi t_L$. Since
\begin{equation}
\frac{t_\phi}{t_L} = 12.4 \left(\frac{10^{-10}{\rm eV}}{m_\phi}\right)
\left(\frac{1 \,{\rm km}}{L}\right)\,,
\label{eq:tltphi}
\end{equation}
we obtain
\begin{equation}
m_{\phi, \rm Compton}^L = 2 \times 10^{-10} {\rm eV}\left(\frac{1 {\rm km}}{L}\right).
\label{eq:mcritL}
\end{equation}

Let us compute the ensemble-averaged oscillation probabilities in Eq.~\eqref{eq:ensembleav}
in different ranges of mass $m_\phi$ as summarised in Table~\ref{tab:regimes_real}. 
The probability for an individual neutrino $P(t_0, t_L)$ Eq.~\eqref{eq:prob-osc}
has the oscillation phase given by Eq.~\eqref{eq:phase-osc} with 
$\Delta m^2_{\rm ref}(\mathbf{x},t)$ from Eq.~\eqref{eq:modulation}. Explicitly, 
\begin{equation}
\phi_{\rm osc}= \int_{t_0}^{t_0 + t_L}dt \frac{\overline{\Delta m^2_{\rm ref}}(\mathbf{x},t)}{2E}
\left[1+\cos(2m_\phi t + \vartheta)\right]. 
\label{eq:phasereal}
\end{equation}

\begin{enumerate}[label=\Roman*.]
\item If $t_\phi \ll t_{\rm coh} \ll t_L$ (similar to the case I for complex field), both the Compton and the pattern variations should be taken into account at the level of individual neutrino propagation. 
This leads to averaging of both fast oscillations and variations of 
$\Delta m^2_{\rm ref}$ in Eq.~\eqref{eq:phasereal}: 
\begin{equation}
\phi_{\rm osc}= \frac{\langle \overline{\Delta m^2}(t_0, t_L) \rangle t_L}{2E}.
\label{eq:phasereal1}
\end{equation}
where 
$$
\langle \overline{\Delta m^2}(t_0, t_L) \rangle = 
\frac{1}{t_L}\int_{t_0}^{t_0 + t_L}dt \overline{\Delta m^2_{\rm ref}}(\mathbf{x},t) \left[1+\cos(2m_\phi t + \vartheta)\right]
$$
Ensemble averaging does not change the result. Also, the dependence on $t_0$ is negligible. 
Such a situation is realised for $m_\phi > 10^{-2}$ eV (see Fig.~\ref{fig:scales} line a and regime a in Table~\ref{tab:regimes_real}). 
There is no distortion of the probability and hence, no bound on the refractive mass in this case.

\item If $t_{\rm coh} \gg t_L$, the slowly changing factor due to pattern variations can be put out of the integral 
Eq.~\eqref{eq:phasereal} and the phase becomes 
\begin{equation}
\phi_{\rm osc}= \frac{\overline{\Delta m^2}(t_0)}{2E}
\int_{t_0}^{t_0 + t_L}dt \left[1+\cos(2m_\phi t + \vartheta)\right].
\label{eq:phasereal2}
\end{equation}
After integration, we get
\begin{equation}
\phi_{\rm osc}= \frac{\overline{\Delta m^2}(t_0)}{2E}
\left\{1 + K\cos\left[\frac{4\pi}{t_\phi} \left(t_0 + \frac{t_L}{2}\right)+\vartheta\right]\right\},\qquad 
K=\frac{\sin(2\pi t_L/t_\phi)}{2\pi t_L/t_\phi},
\label{eq:phasereal3}
\end{equation}
where $m_\phi$ has been substituted by $2\pi/ t_\phi$.

There are two cases here:

(i) If $2\pi t_L \gg t_\phi$, then $K \ll1$ and the oscillatory term in $\phi_{\rm osc}$ 
can be neglected, so that the phase becomes
\begin{equation}
\phi_{\rm osc}= \frac{\overline{\Delta m^2}(t_0)}{2E}.
\label{eq:phasereal4}
\end{equation}
It corresponds to the averaging of fast oscillations in $\Delta m^2_{\rm ref}$ (see line b in Fig.~\ref{fig:scales} and regime b in Table~\ref{tab:regimes_real}).
The pattern variation should be accounted 
for at the ensemble level as in complex field case II. 
This is realised for the range $m_\phi = [10^{-9}, 10^{-8}]$ eV for temporal conditions, 
but never realised if spatial conditions are taken into account. This is because $\lambda_{\rm dB}> L$ provides a stronger constraint and does not allow for the above range of $m_\phi$. 

(ii) If $2\pi t_L \ll t_\phi$, then $K \rightarrow 1$ and 
\begin{equation}
\phi_{\rm osc}= \frac{\overline{\Delta m^2}(t_0)}{2E}
\left\{1 + \cos\left[\frac{4\pi}{t_\phi} \left(t_0 + \frac{t_L}{2}\right)+\vartheta\right]\right\}. 
\label{eq:phasereal5}
\end{equation}
For $t_\phi = [t_L^{\rm min},t_L^{\rm max}]$, we have $t_{\rm coh} \gg t_L^{\rm max}c/\bar{v}$. 
The Compton period lies in the transition region (see line c in Fig.~\ref{fig:scales}), where the appropriate averaging depends 
on the experimental baseline. Experiments with long baselines average the Compton modulation along each neutrino path. In contrast, those with short baselines only need to consider 
the ensemble averaging, and the pattern variations should be accounted for at the probability level. 
Such a situation is realised for $m_\phi = [10^{-13} , 10^{-10}]$ eV.

Let us consider the averaging of probability over an ensemble of events 
detected during $t_{\rm bin}$. From Eq.~\eqref{eq:ensembleav},
\begin{equation}
\langle P(t_0, t_L) \rangle = 1 - \frac{\sin^2 2\theta}{2} \left[1 - 
\frac{1}{t_{\rm bin}} \int_{t_{\rm in}}^{t_{\rm in} + t_{\rm bin}} dt_0 \cos \phi_{\rm osc}(t_0, t_L) \right] 
\label{eq:ensemble22}
\end{equation}
and the phase is given in Eq.~\eqref{eq:phasereal5}. 
The integral can be computed in two steps. We first divide whole integration time $t_{\rm bin}$ into intervals of
$\Delta t_0 = \pi t_\phi$ which are much smaller than $t_{\rm coh}$. Within each of these intervals, for each point in ($\tilde{t}_0,\tilde{t}_0 + \Delta t_0$), we have
$\overline{\Delta m^2}(t_0) = \overline{\Delta m^2}(\tilde{t}_0)$ is constant. 
Integration over intervals $\Delta t_0$ (averaging of probability) gives 
\begin{equation}
P(\tilde{t}_0) = 
1-\frac{1}{2}\sin^2 2\theta \left[ 1-J_0\left(\frac{K L\overline{\Delta m^2}(\tilde{t}_0)}{2E}\right) 
\cos\left(\frac{\overline{\Delta m^2}(\tilde{t}_0)L}{2E}\right)\right],
\label{eq:J0}
\end{equation}
where $J_0$ is the zeroth-order Bessel function. 
Secondly, we perform averaging over $\tilde{t}_0$: 
\begin{equation}
\langle P(\tilde{t}_0) \rangle = \frac{1}{t_{\rm bin}} \int_{t_{\rm in}}^{t_{\rm in} + t_{\rm bin}} d\tilde{t}_0 P(\tilde{t}_0). 
\label{eq:pav}
\end{equation}
For $t_{\rm bin}\gg t_{\rm coh}$, the detector samples many independent 
density configurations, and the time average in Eq.~\eqref{eq:pav} reduces to an average over the density distribution:
\begin{equation}
\langle P\rangle_{t_{\rm bin}} =\int d\rho \, R(\rho)\, P(\overline{\Delta m^2}(\rho)). 
\label{eq:density_average}
\end{equation}
Here we have used the mapping of $\overline{\Delta m^2}(\tilde{t}_0)\to \overline{\Delta m^2}(\rho)$ described 
in Eqs. (\ref{eq:int-rho}) - (\ref{eq:int-rho2}).
In Eq. (\ref{eq:density_average}) $R(\rho)$ is the distribution of the slowly-changing field density, 
$\rho = |\psi|^2$.

\begin{table}[t]
\caption{Regimes of field variations for a real scalar field, in order of
decreasing $m_\phi$. In addition to the interference pattern, the field squared
oscillates with the Compton period $t_\phi$. Notation as in
Table~\ref{tab:regimes_complex}. The boundaries follow from
$t_{\rm coh}=t_L$, $t_\phi=2\pi t_L$, $t_{\rm coh}=t_{\rm bin}$ and
$t_\phi=t_{\rm bin}$. Since region c in Fig.~\ref{fig:scales} represents a transition regime, we do not include it in the table.}
\label{tab:regimes_real}
\begin{ruledtabular}
\begin{tabular}{clllc}
Regime & Timescale condition & \begin{tabular}[b]{@{}l@{}}Interference\\ pattern\end{tabular} & \begin{tabular}[b]{@{}l@{}}Compton\\ oscillations\end{tabular} & $m_\phi$ (eV) \\
\hline
a & $t_\phi\ll t_{\rm coh}\ll t_L\ll t_{\rm bin}$ & path & path   & $\gtrsim10^{-3}\,(1\,{\rm km}/L)$ \\
b & $t_\phi\ll 2\pi t_L\ll t_{\rm coh}\ll t_{\rm bin}$ & ensemble & path & $10^{-10}\,(1\,{\rm km}/L)$--$10^{-3}\,(1\,{\rm km}/L)$ \\
d & $2\pi t_L\ll t_\phi\ll t_{\rm coh}\ll t_{\rm bin}$ & ensemble & ensemble & $10^{-16}\,(1\,{\rm yr}/t_{\rm bin})$--$10^{-10}\,(1\,{\rm km}/L)$\\
e & $2\pi t_L\ll t_\phi\ll t_{\rm bin}\ll t_{\rm coh}$      & frozen          & ensemble   & $10^{-22}\,(1\,{\rm yr}/t_{\rm bin})$--$10^{-16}\,(1\,{\rm yr}/t_{\rm bin})$ \\
f & $t_L\ll t_{\rm bin}\ll t_\phi\ll t_{\rm coh}$ & frozen & frozen & $ \lesssim 10^{-22}\,(1\,{\rm yr}/t_{\rm bin})$ \\
\end{tabular}
\end{ruledtabular}
\end{table}

\begin{figure}
 \centering
 \includegraphics[width= 1. \linewidth]{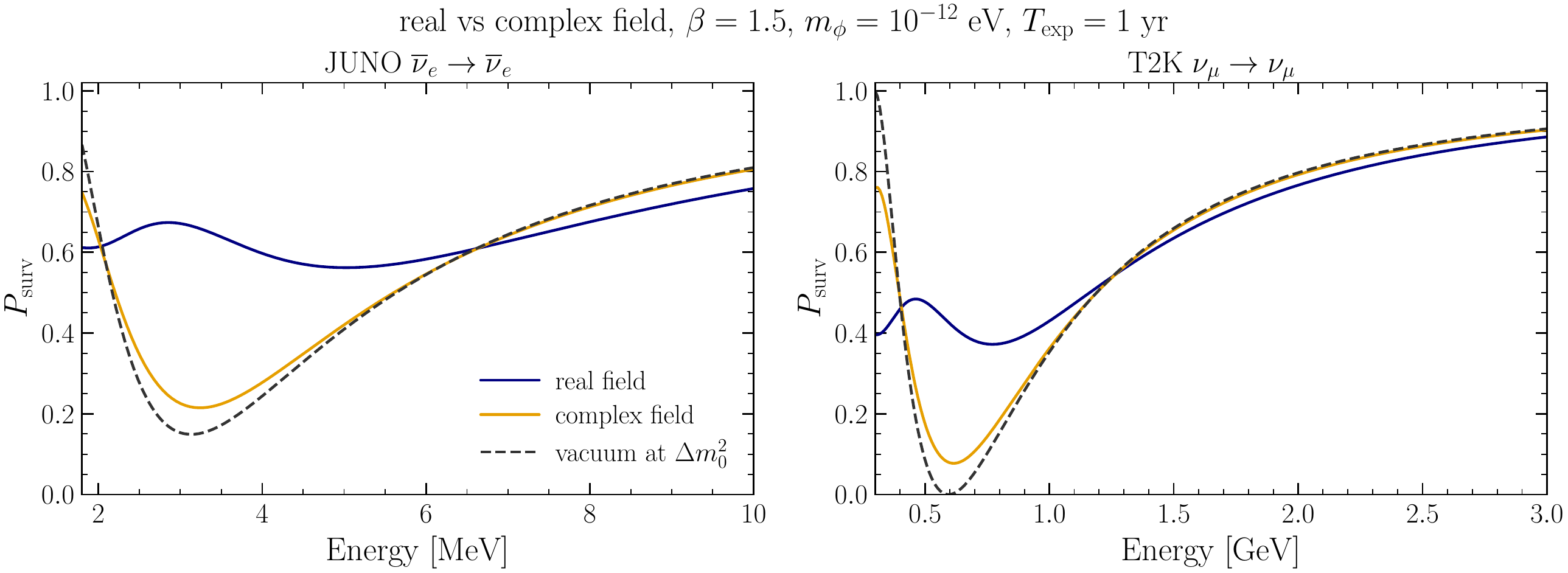}
 \caption{Comparison of neutrino oscillation probabilities for real and complex scalar fields 
at $m_\phi = 10^{-12}~\mathrm{eV}$ and $\beta = 1.5$.}
 \label{real}
\end{figure}
\end{enumerate}

Fig.~\ref{real} compares the neutrino oscillation probabilities for real and complex scalar fields, 
for the benchmark values $m_\phi=10^{-12}~\mathrm{eV}$ and $\beta=1.5$. While the self-interactions 
restore visible oscillatory features in the complex-field case, as shown in Fig.~\ref{prob12}, 
the Bessel function modulation of the real field washes out these features. 
Consequently, the oscillations remain suppressed despite the presence of self-interactions, in the case of a real field.
This consideration corresponds to line d in Fig.~\ref{fig:scales} and Table~\ref{tab:regimes_real}.

With a further decrease of $m_\phi$, two other cases are realised.
If $t_\phi \ll t_{\rm bin}$ and $t_{\rm coh} \gg t_{\rm bin}$ (line e in Fig.~\ref{fig:scales} and regime e in Table~\ref{tab:regimes_real}), the Compton variations should be accounted for at the ensemble level while the pattern variations are frozen. This happens when $m_\phi \ll 10^{-16}$ eV. In this case, the probability is given by Eq.~\eqref{eq:J0}, and no further averaging over $\tilde{t}_0$ is needed. Now $\overline{\Delta m^2}(\tilde{t}_0)$ is constant and should be fixed by tuning with data. Still, there is a distortion of the probability due to the presence of the Bessel function. This excludes the mass range $m_\phi=[10^{-22},10^{-16}]\,$eV.

In the presence of this Compton modulation, the condition that determines the critical mass is 
\begin{equation}
t_{\rm bin} = t_{\phi}\,. 
\label{eq:tt1}
\end{equation}
It gives
\begin{equation}
 m_\phi^{\rm cr} = \frac{2\pi}{t_{\rm bin}} = 10^{-22}{\rm eV} 
\left(\frac{1\,{\rm year}}{t_{\rm bin}}\right). 
\label{eq:mcritr}
\end{equation}
This is about six orders of magnitude smaller than the critical mass of the interference pattern $(m_\phi^{\rm cr})$ in Eq.~\eqref{eq:mcrit}.

The condition (\ref{eq:tt1}) determines another extreme case: $t_\phi \gg t_{\rm bin}$ when all variations are frozen during experimental time. This happens for $m_\phi \ll 10^{-22}$ eV (See line f in Fig.~\ref{fig:scales} and regime f in Table~\ref{tab:regimes_real}).
Thus, apart from extreme cases for a real field, the oscillations 
are strongly affected over the whole mass range of interest.
An increase in $\beta$ does not change this conclusion.

For a real field, 
the additional Compton modulation makes the short-baseline reactor experiments 
Daya Bay, RENO and Double Chooz particularly relevant, 
since they retain the full modulation during propagation up to 
$m_\phi\sim10^{-10}$~eV. Solar and supernova neutrinos are less sensitive to these effects because 
they do not resolve the oscillatory pattern in $L/E$.

\section{Effects of modified interference pattern}
\label{sec:modif}

In what follows, we will consider further possible modifications of the interference 
patterns and their effects on oscillations. 
We will focus on modifications that allow us to relax or completely 
remove the bounds on the refractive neutrino mass. 
Based on the consideration in the previous section, such a modification can be made in three directions: 

\begin{enumerate}

 \item Increase of the scale of the interference pattern, that is, the size of the granules and voids. The increase can be due to a reduction in the average virial velocity of scalars. For this, new physics and new scales should be involved, thereby changing the dynamics of field formation.
 \item Decrease in the amplitude of variations of density around the average value. That would lead to a narrowing of the density distribution and, consequently, to a narrowing of $R(\rho) = R(\Delta m_{\rm ref}^2)$. The effect of probability averaging is reduced. In the extreme case: $R(\Delta m_{\rm ref}^2) \rightarrow \delta(\Delta m_{\rm ref}^2 - \overline{\Delta m^2_{\rm ref}})$. 
 \item Averaging of the density distribution due to the effect of several scalars, overlap of granules, etc.
\end{enumerate} 
The following factors can modify the interference pattern and, consequently, change the $\Delta m_{\rm ref}^2$ distribution:
\begin{itemize}
\item Interaction with the gravitational field of the Sun and the Earth.
\item Effect of many scalars. 
\item Interactions with cosmic backgrounds: relic neutrinos (C$\nu$B), photons (CMB), gravitational waves.
\item Damping of fluctuations, thermalisation.
\item Effects of the DM production mechanisms, virialisation, and relaxation. 
\item Production of big-size bosonic (axion) stars with small velocity inside.
\end{itemize}
We will consider the first three items in some detail.
 
\subsection{Effects of gravitational field of the Sun and the Earth}
The presence of external gravitational potential from the Sun and the Earth can be accounted 
for by substitution in the evolution equation Eq.~\eqref{eq:tildeGPP} 
\begin{equation}
 \Phi \rightarrow \left(\Phi - \frac{GM_\odot }{r}\right)
\label{eq:externg1}
\end{equation}
where $r = \sqrt{x^2 + y^2 + z^2}$. 
Under the rescaling, $r = \tilde{r}/m_\phi \bar{v} $, 
the equations become 
\begin{align}
i \partial_{\tilde{t}} \tilde{\psi} & = -\frac{1}{2}\tilde{\nabla}^2 \tilde{\psi} + 
\left(\tilde{\Phi} - \frac{GM_\odot m_\phi}{\tilde{r}\bar v}\right) \tilde{\psi} 
+ \beta|\tilde{\psi}|^2 \tilde{\psi}, \\
\tilde{\nabla}^2 \tilde{\Phi} & = \alpha(|\tilde{\psi}|^2 - 1). 
\label{eq:externg2}
\end{align}

In our simulations, the GPP equations are solved with periodic boundary conditions, which are not well-suited to modelling an isolated massive body like the Sun and its gravitational focusing effect. A consistent treatment would require an approach similar to a scattering analysis of an interacting ULDM halo in the Sun's external gravitational potential. This is left for future work; here we limit ourselves to a qualitative discussion based on the intuition provided by previous analyses of gravitational focusing in the absence of self-interactions.

The effect of the external gravitational field is strong if the size 
of the object (or region of strong field), $r_g$, is bigger or
comparable to the granular size, $\lambda_{\rm dB}$. Applications of this effect have been considered in \cite{Gherghetta:2023myo}.

For the Sun with $R_{\rm sun} = 7 \times 10^{5}$ km, the condition
$2R_{\rm sun} \geq \lambda _{\rm dB}$ is satisfied for
$m_\phi > 1.24 \times 10^{-12}$ eV.
So, for $m_\phi \ll 10^{-12}$ eV, the distortion of the probability is small. The Earth effect is strong if 
$m_\phi > 1.5 \times 10^{-11}$ eV.

The external gravitational field can lead to the distortion
and disruption of granules. For this, one should compare the gravitational force 
of the Sun with the gravitational force produced by the granule itself. The mass of granules equals 
$$
\begin{aligned}
M_G &\equiv B\,\rho_{\rm DM}V_G
\simeq 4B\,\rho_{\rm DM}\lambda_{\rm dB}^3\\
&\simeq 1.3\times10^{25}\,{\rm GeV} \,B
\left(\frac{10^{-10}\,{\rm eV}}{m_\phi}\right)^3\\
&\simeq 20\,{\rm g}\,B
\left(\frac{10^{-10}\,{\rm eV}}{m_\phi}\right)^3.
\end{aligned}
$$
Here, $B = O(10)$ denotes the DM overdensity in the granule.
Let us consider a particle of mass $m$ at the border of the granule
at the Earth orbit, $R_{\rm orb} \approx 1.5 \times 10^{8}$ km. 
The gravitational force of the Sun on this particle
equals $F_\odot = G_N M_\odot m/R_{\rm orb}^2$, while the gravitational 
force of granule itself is $F_{\rm gran} = G_N M_G m/\lambda_{\rm dB}^2$. 
The ratio of the two forces: 
\begin{equation}
\frac{F_\odot}{F_{\rm gran}} = \frac{M_\odot}{M_G}
\left(\frac{\lambda_{\rm dB}}{R_{\rm orb}} \right)^2 = \frac{M_\odot}{4B\rho_{\rm DM}}\frac{1}{\lambda_{\rm dB}R_{\rm orb}^2}. 
\label{eq:f-ratio}
\end{equation}
For $m_\phi= 10^{-10}$ eV the ratio equals $2 \times 10^{22}$. 
The gradient of the solar force over the granular gravitational 
force ($\Delta F_\odot = \lambda_{\rm dB} dF_\odot /dr $): 
\begin{equation}
\frac{\Delta F_\odot}{F_{\rm gran}} = 2\frac{M_\odot}{M_G}
\left(\frac{\lambda_{\rm dB}}{R_{\rm orb}} \right)^3
\label{eq:f-ratio2}
\end{equation}
equals $5\times 10^{17}$. So, the solar gravity is much stronger than 
the self-gravity, which holds the granules together. Therefore, the solar force might modify and destroy the granules. That would lead to a smoother density distribution, 
and consequently, to the reduction of the averaging effect. 
According to Eq.~\eqref{eq:f-ratio}
$$
\frac{F_\odot}{F_{\rm gran}} \propto 
1/\lambda_{\rm dB} \propto m_\phi. 
$$
So, with a decrease in $m_\phi$, the ratio of forces decreases, 
and the gravitational effect of the Sun becomes less dominant. 
However, the relative gradient does not change. The consideration does not work when the size of 
the granule becomes bigger than the Earth's orbit.

Following Ref.\cite{Kim_2022}, which studies the gravitational focusing of ULDM by the Sun in the absence of self-interactions, the resulting density enhancement is typically at the few percent level for a smooth galactic halo. 
The magnitude of the effect depends on the de Broglie wavelength: for sufficiently large masses, the wave solution approaches the classical particle-focusing limit, yielding an overdensity 
of a few percent near the Sun, whereas for lower masses, where the de Broglie wavelength exceeds the relevant scales in the solar neighbourhood, the effect is strongly suppressed. This effect is much stronger for DM streams and dark disks, as discussed in Ref.~\cite{Kim_2022}. As a preliminary estimate, however, we therefore expect their impact on the averaged density distribution, and hence on the oscillation probabilities, to remain subdominant. A more complete treatment, particularly in the presence of self-interactions, is left for future work.

\subsection{Effect of many scalar fields}

The density distribution $R$ changes if several scalars 
$\phi_i$ with different masses $m_{\phi,i}$ contribute to the interference pattern. 
The axiverse scenario supports such a possibility \cite{Arvanitaki:2009fg}, according to 
which up to several dozen axion-like particles may exist. 
In the absence of strong self-interactions, they produce independent interference patterns with different spatial and time 
 scales given by their de Broglie lengths and the coherence times, which are inversely proportional to $m_{\phi, i}$. 
The overlap of these patterns averages density fluctuations and, therefore, density spread, narrowing the distributions.

\begin{figure}[!t]
 \centering
 \includegraphics[width=1.0 \linewidth]{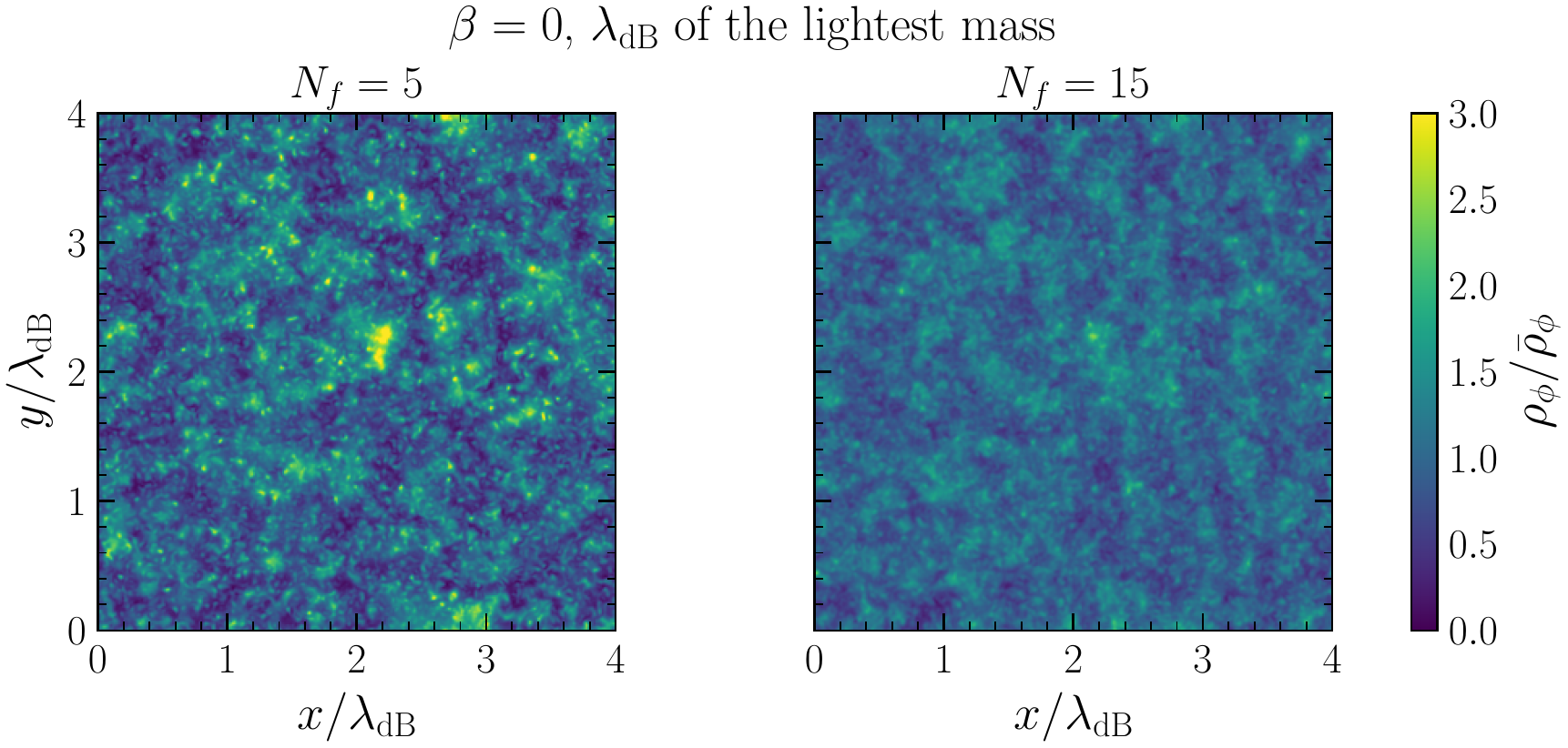}
\caption{Comparison of the interference patterns produced by 5 and 15 scalar fields in the absence of self-interactions ($\beta=0$). The de Broglie wavelength of the lightest field sets the box size.}
\label{pattern15s}
\end{figure}
\begin{figure}[!t]
 \centering
 \includegraphics[width=0.6 \linewidth]{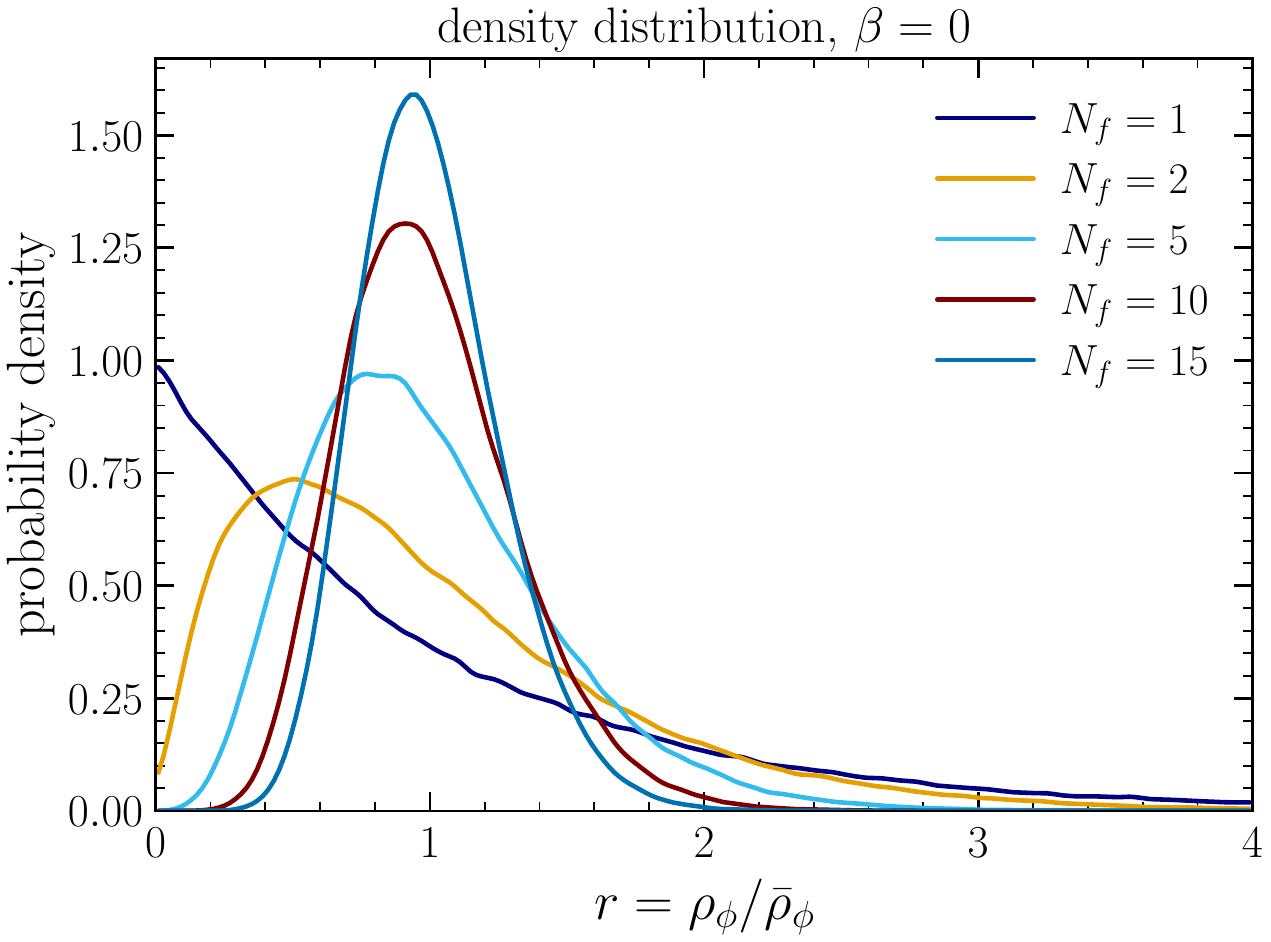}
\caption{Density distributions for different numbers of scalar fields, with their masses logarithmically spaced over the range $10^{-12}$--$10^{-11}$~eV.}
\label{density15s}
\end{figure}
\begin{figure}
 \centering
 \includegraphics[width=1.0 \linewidth]{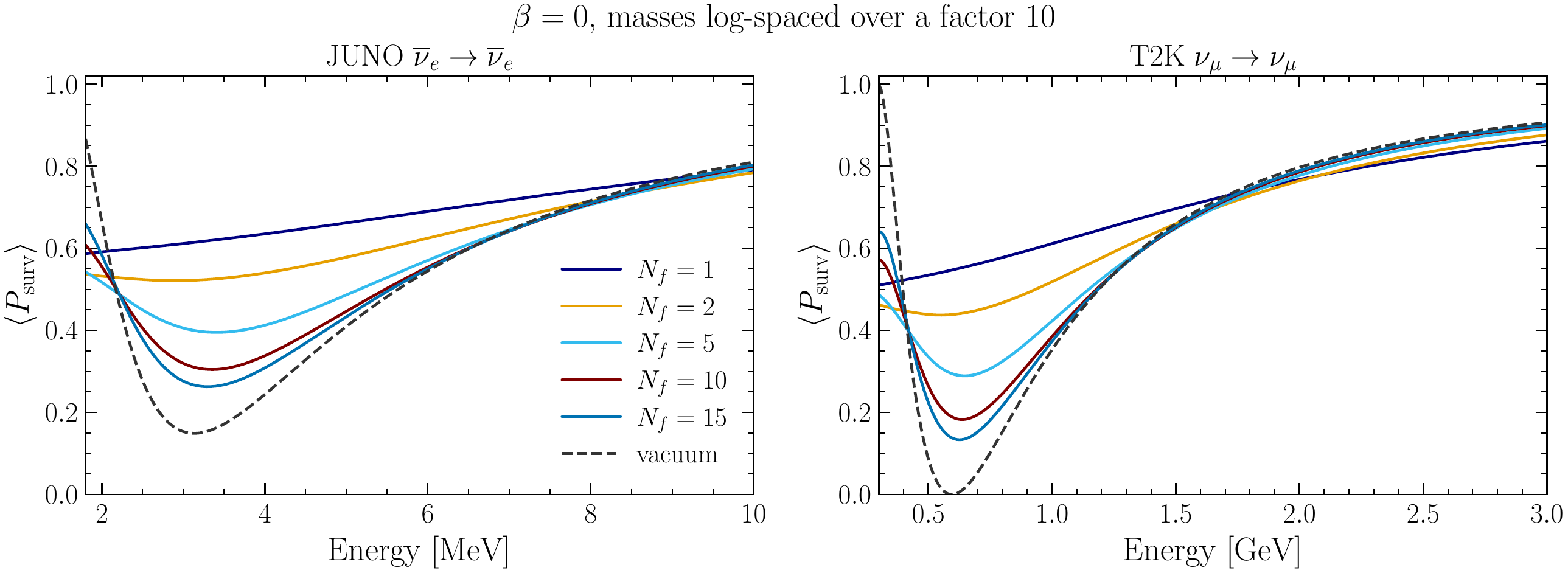}
\caption{Oscillation probability for different numbers of scalar fields, with their masses logarithmically spaced over the range $m_\phi \in [10^{-12},10^{-11}]$~eV.}
\label{probability15s}
\end{figure}

In Figs.~\ref{pattern15s}, \ref{density15s} and \ref{probability15s} 
we show the interference patterns, the density distributions 
and the probabilities for different numbers of scalars. We choose logarithmically spaced masses for the scalar species. While each field can be rescaled according to its mass, the same rescaling does not apply to the total field, since the different species 
have distinct coherence lengths and coherence times, resulting in mutually incoherent granule patterns.

As follows from Fig.~\ref{pattern15s}, with an increase in $N_f$, the pattern becomes more uniform, with suppression of large and small density regions. Correspondingly, the density distributions (Fig.~\ref{density15s}) become narrower, and maxima shift to larger $\rho$. Consequently, the averaging (distortion) of probabilities weakens. 

Extrapolating results of Fig.~\ref{probability15s} to larger $N_f$
we find that in the oscillation minimum the change of the probability 
by $\Delta P = 0.1$ can be obtained at $N_f = 17-20$; the change 
$\Delta P = 0.05$ is achieved at $N_f > 30$. The averaging effect increases with decreasing energy; however, at low energies in the experiments, background also increases, and energy resolution worsens.

The presence of multiple scalar fields also suppresses the Compton-frequency modulation in the case of real scalars, potentially weakening or removing the corresponding bounds on refraction-induced neutrino masses. Since the Compton phases of different scalar species are uncorrelated, their modulations partially cancel, reducing the amplitude of the total modulation as $1/\sqrt{N_f}$. Consequently, the argument of the Bessel function governing the oscillation probability is also suppressed as $1/\sqrt{N_f}$, so that the resulting distortion scales as $1/N_f$ in the small-modulation limit. A sufficiently large number of scalar species can therefore substantially reduce the Compton-induced distortion and relax the constraints on the refractive mass scenario. 

\subsection{Other possible modifications}

Interactions between the scalar field and the thermal 
neutrino bath can generate thermal corrections to the scalar effective potential 
and mass, as well as dissipative effects~\cite{Banerjee:2025nvs, Banerjee:2025qyg}. 
The oscillating classical scalar field continuously modifies 
the effective neutrino masses and, consequently, their instantaneous 
distribution function. Since the neutrino bath takes a finite time to thermalise with the scalar background, 
there is a resultant energy transfer from the scalar field to the neutrinos, causing an effective damping.

However, note that the results of Refs.~\cite{Banerjee:2025nvs, Banerjee:2025qyg} assume that the neutrino masses arise from a direct coupling with the scalar field with no extra mediator. In our model, described by Eq.~(\ref{eq:lagran}), the mass depends on $\phi$ only through $\nu$-$\chi$ mixing, with the additional light fermion $\chi$ also participating in the dynamics. Whether analogous thermal corrections or dissipative effects arise therefore requires a dedicated calculation, which is beyond the scope of this work. Such effects would primarily modify the cosmological relic abundance and hence the coupling required in Eq.~\eqref{eq:strength}; their possible impact on subsequent halo formation is not considered here. For the present-day halo, the relic neutrino background lies below the required resonance energy, giving a negligible scattering rate. Couplings to present-day photon and gravitational-wave backgrounds arise only through higher-order loops and gravity, respectively, and are also negligible. We therefore neglect interactions with present-day cosmic backgrounds in determining $R(\rho)$.

Another possibility for modifying the interference pattern is a smaller velocity dispersion $\bar{v}$. For a fixed scalar mass $m_\phi$, it reduces the characteristic momentum of the field and increases both the de Broglie wavelength and the coherence time. Near the critical mass, $\lambda_{\rm dB}$ already exceeds every baseline, so the relevant condition is $t_{\rm coh}\gg t_{\rm exp}$. The density pattern is then frozen over the exposure, and the averaging over $R(\rho)$ that would otherwise distort the oscillation spectrum is absent. Consequently, the corresponding bounds on the refractive mass for a complex scalar field disappear. Since $t_{\rm coh}=2\pi/(m_\phi\bar{v}^2)$, the critical mass scales as $\bar{v}^{-2}$, so a smaller velocity dispersion extends the
allowed range to larger $m_\phi$. The Compton-frequency modulation of a real scalar field, however, is independent of $\bar{v}$ and is therefore unaffected by this mechanism.

The extreme case $\bar{v}\rightarrow 0$ corresponds to a fully coherent bosonic condensate, for which $\lambda_{\rm dB}\rightarrow\infty$ and $t_{\rm coh}\rightarrow\infty$. In a virialised galactic halo, however, the velocity dispersion is set by the gravitational potential through the virial theorem. A substantially smaller $\bar{v}$ would therefore require a local field configuration with a smaller internal velocity dispersion,
such as a boson star of sufficiently large spatial extent, or a DM field produced through a formation mechanism that leaves it non-virialised.

\section{Conclusion}
\label{sec:conclusion}

We have investigated the effects of spatial and temporal fluctuations of ULDM with quartic self-interactions on refractive neutrino masses and oscillation probabilities. 
We solve the Gross-Pitaevskii-Poisson (GPP) equations describing 
the evolution of the scalar field and exploit scaling properties of the equations to obtain solutions over a wide range of scalar masses. 
We simulate the interference patterns of the field for different self-interaction strengths 
and determine the density distributions sampled during neutrino propagation and experimental exposure. 
These distributions govern the averaging of the oscillation probabilities and, consequently, 
the extent to which the energy dependence of the probability is distorted.

Self-interactions modify the interference pattern in two ways: repulsive interactions smooth the density contrast, narrowing $R(\rho)$ and thereby weakening the averaging of oscillation probabilities, while interactions of either sign reduce the granule size and coherence time. The latter effect is particularly pronounced in the coherence time, indicating that stronger self-interactions produce smaller, more rapidly evolving density structures.

For a complex scalar field, only fluctuations of the interference pattern contribute to the variation 
of the refractive mass. For $m_\phi\lesssim10^{-16}$ eV, the pattern remains effectively frozen during 
the experimental exposure, and its effect can be absorbed into a rescaling of $\Delta m^2$, 
leaving the refractive mass scenario viable against these constraints. At larger masses, averaging over 
the density distribution distorts the energy dependence of the oscillation probabilities. 
In the absence of self-interactions, this effect strongly disfavours the mass range $m_\phi \sim [10^{-15},10^{-9}]$ eV. Repulsive self-interactions weaken the averaging by narrowing the normalised density distribution and shifting its maximum towards the mean density. For $\beta=1.5$, the deviation of the survival probability at the first oscillation minimum is reduced to $\Delta P\simeq0.07$ in both JUNO and T2K. Extrapolating to $\beta\simeq2.2-2.4$ reduces the deviation further to $\Delta P\simeq0.05$, potentially relaxing the corresponding constraints for $m_\phi \gtrsim 10^{-15}$ eV.

For a real scalar field, the density also oscillates at the Compton frequency $2m_\phi$, introducing an additional source of distortion described by the Bessel-function modulation in Eq.~\eqref{eq:J0}. For $10^{-22}\lesssim m_\phi\lesssim10^{-10}$ eV, these rapid temporal variations distort the oscillation probabilities. Unlike the distortions arising from the interference pattern, this Compton-induced effect cannot be eliminated by repulsive self-interactions. For $m_\phi\ll10^{-22}$ eV, the Compton modulation is effectively frozen during the experimental exposure, while for $m_\phi\gtrsim10^{-9}$ eV, the rapid oscillations are averaged along individual neutrino trajectories and their additional distortion becomes negligible. Thus, the real-field scenario is subject to additional constraints over a substantially wider mass range than the complex-field case.

We have also explored several possibilities that could modify the interference pattern and weaken the resulting constraints. In particular, the presence of multiple scalar fields $(N_f)$ with different masses produces overlapping interference patterns with distinct spatial and temporal scales, smoothing the total density distribution and reducing the averaging effect. We find that the deviation in the oscillation probability can be reduced to $\Delta P\simeq0.1$ for $N_f=17$-$20$ and decreases further with increasing $N_f$. For real scalar fields, the incoherent addition of Compton phases also suppresses the modulation amplitude, providing an additional mechanism for relaxing the constraints. The gravitational fields of the Sun and the Earth may also modify the granule structure when the de Broglie wavelength becomes comparable to their characteristic scales. While these effects
could distort or disrupt individual granules, estimates of
gravitational focusing for a smooth halo suggest only a
few percent enhancement of the local density. We therefore expect
their impact on the averaged density distribution, and hence on
the oscillation probabilities, to remain subdominant. We have also discussed possible effects of interactions with the cosmic neutrino background in the early Universe and the formation of bosonic structures with small internal velocity dispersion, which could further modify the field configuration.

Overall, our results demonstrate that the viability of refractive neutrino masses depends crucially on the nature of the scalar field and its interference pattern. A broad range of scalar masses 
remains viable in the complex-field scenario, with repulsive self-interactions as well as with multiple scalar species potentially extending the allowed region of $m_\phi$. In contrast, real scalar fields are subject to additional constraints from Compton-frequency modulation, which disfavour an extended mass range.

\section*{Acknowledgments}

The authors are grateful to Giovanni Villadoro, Mehrdad Mirbabayi, and David Cyncynates 
for very useful discussions. MS acknowledges support from 
the Early Career Research Grant by Anusandhan National Research Foundation (ANRF/ECRG/2024/000522/PMS). 
The work of MS is supported by the Max-Planck-Gesellschaft through a Max Planck-Partner group. 
\bibliography{references}
\newpage
\appendix
\section{Numerical method}
\label{app:numerics}
We evolve the field in discrete time steps $\Delta\tilde t$ using the pseudo-spectral split-step
Fourier method~\cite{Woo_2009, Edwards_2018}. We write the Hamiltonian in Eq.~\eqref{eq:tildeGPP} as
$\hat H=\hat T+V$, with the kinetic operator $\hat T=-\tfrac12\tilde\nabla^2$ and the potential
$V=\tilde\Phi+\beta|\tilde\psi|^2$. Since $\hat T$ and $V$ do not commute, one step is approximated by
the symmetric (Strang) product~\cite{strang1968},
\begin{equation}
e^{-i\hat H\Delta\tilde t}=e^{-iV\Delta\tilde t/2}\,e^{-i\hat T\Delta\tilde t}\,e^{-iV\Delta\tilde t/2}
+O(\Delta\tilde t^3),
\label{eq:strangop}
\end{equation}
i.e., three successive operations,
\begin{equation}
\begin{aligned}
\tilde\psi_1 &= e^{-iV\Delta\tilde t/2}\,\tilde\psi(\tilde t),\\
\tilde\psi_2 &= \mathcal F^{-1}\!\left[e^{-i\tilde k^2\Delta\tilde t/2}\,\mathcal F[\tilde\psi_1]\right],\\
\tilde\psi(\tilde t+\Delta\tilde t) &= e^{-iV_2\Delta\tilde t/2}\,\tilde\psi_2,
\end{aligned}
\label{eq:strang}
\end{equation}
where $\mathcal F$ is the discrete Fourier transform and $V_2$ is the potential recomputed from
$\tilde\psi_2$. The potential steps are local phase rotations in position space; the kinetic step
is exact in Fourier space. The gravitational potential is obtained by solving the Poisson
equation in Fourier space with the zero mode removed. Each operation is unitary, so the total
mass $\int d^3\tilde x\,|\tilde\psi|^2$ is conserved to numerical precision, and the symmetric
ordering makes the scheme second-order accurate in time.

\textit{Resolution.} The grid must resolve the shortest length scale of the field. For $\beta=0$
this is the granule size, set by the velocity dispersion. For repulsive self-interactions, a
shorter scale appears, the healing length $\zeta$: the length over which a density variation costs
as much kinetic (gradient) energy as the self-interaction energy at the mean density. Equating the
two in rescaled units, $1/(2\tilde\zeta^{2})=\beta$, gives
\begin{equation}
\zeta=\frac{\lambda_{\rm dB}}{2\pi\sqrt{2\beta}}.
\label{eq:healing}
\end{equation}
On scales smaller than $\zeta$, the kinetic term dominates, and the field behaves as a free wave, as
for $\beta=0$. On larger scales, the self-interaction dominates, and density perturbations
propagate as sound waves on the uniform background. Increasing the repulsive coupling therefore
decreases $\zeta$ and moves the transition between the two regimes to smaller scales. A grid
spacing $\Delta\tilde x\gtrsim\tilde\zeta$ would cut off the free-wave part of the excitation
spectrum and misrepresent the relaxation, so we require $\Delta x\ll\zeta$ for every $\beta$
considered. We use a periodic box of side $L_{\rm box}=4\lambda_{\rm dB}$ on a $256^3$ grid, so $\Delta x=\lambda_{\rm dB}/64$. 

\textit{Time step.} The step resolves both the kinetic evolution and the phase accumulated
through the self-interaction:
\begin{equation}
\Delta t=\min\left(\frac{\Delta\tilde x^2}{2\pi},\ \frac{0.2}{|\beta|\,\tilde\rho_{\max}}\right),
\end{equation}
where $\tilde\rho_{\max}$ is the maximum of $|\tilde\psi|^2$ in the initial field. The first condition
limits the kinetic phase per step at the Nyquist wavenumber to $\pi/4$, the second limits the
self-interaction phase to $0.2$. For $\beta<0$ the density can grow during the evolution, so the
step is recomputed after every step from the current maximum density. For $\beta=0$ the kinetic
step is exact for any $\Delta\tilde t$, and the step is set by the sampling, $\Delta t\le t_{\rm coh}/32$. Each field is relaxed for $200\,t_{\rm coh}$, after which it is evolved for a further $40\,t_{\rm coh}$ while its statistics are recorded.

\end{document}